\documentclass[11pt]{article}

\usepackage[margin=1in]{geometry}
\usepackage{amsmath, amssymb, amsfonts}
\usepackage{graphicx}
\usepackage{placeins}
\usepackage{float}

\usepackage{microtype}
\usepackage{booktabs}
\usepackage{array}
\usepackage{caption}
\usepackage{xcolor}
\usepackage{hyperref}
\usepackage{pdfpages}

\hypersetup{
  hidelinks,
  pdftitle={Degree-conditioned anneal offsets reshape certified-optimum sampling for minimum vertex cover},
  pdfauthor={Jiwan Kang and Namshik Han}
}

\newcommand{\makepreprinttitle}{%
\begin{center}
\vspace*{-1.8em}
{\LARGE\bfseries Degree-conditioned anneal offsets reshape certified-optimum sampling for minimum vertex cover\par}
\vspace{1.0em}
{\large Jiwan Kang\(^{1}\) and Namshik Han\(^{1,2,3,4,5,6,7,8,9,*}\)\par}
\vspace{0.45em}
{\small
\begin{minipage}{0.98\textwidth}
\centering
\(^{1}\)Department of Quantum Information, Yonsei University, Seoul, Republic of Korea\par
\(^{2}\)Institute for Convergence Research and Education in Advanced Technology, Yonsei University,\\
Seoul, Republic of Korea\par
\(^{3}\)Department of Nano Biomedical Engineering (NanoBME), Advanced Science Institute, Yonsei University,\\
Seoul, Republic of Korea\par
\(^{4}\)Center for Nanomedicine, Institute for Basic Science (IBS), Seoul, Republic of Korea\par
\(^{5}\)Department of Artificial Intelligence, Yonsei University, Seoul, Republic of Korea\par
\(^{6}\)College of Pharmacy, Yonsei University, Seoul, Republic of Korea\par
\(^{7}\)Milner Therapeutics Institute, University of Cambridge, Cambridge, United Kingdom\par
\(^{8}\)Cambridge Centre for AI in Medicine, University of Cambridge, Cambridge, United Kingdom\par
\(^{9}\)Cambridge Stem Cell Institute, University of Cambridge, Cambridge, United Kingdom\par
\end{minipage}\par
}
\vspace{0.4em}
{\footnotesize Email: \href{mailto:jw_kang@yonsei.ac.kr}{jw\_kang@yonsei.ac.kr}; \(^{*}\)Corresponding author: \href{mailto:namshik.han@yonsei.ac.kr}{namshik.han@yonsei.ac.kr}\par}
\end{center}
\vspace{-0.4em}
}
\begin{document}

\makepreprinttitle

\begin{abstract}
Quantum annealers can sample degenerate optima unevenly, so the composition of finite-read output matters when multiple solutions are sought. We investigate whether problem structure can steer this output through node-dependent anneal offsets for Minimum Vertex Cover (MVC). In MVC, vertex degree equals the number of incident edge-cover constraints and therefore measures local constraint participation. We use this relation to define degree-conditioned offset-scheduled quantum annealing (DC-OSQA), a static rule assigning more negative offsets to vertices with larger max-normalized degree. Experiments used a D-Wave \texttt{Advantage2\_system1} quantum processing unit (QPU). For each graph, DC-OSQA was compared with a zero-offset baseline using the same MVC encoding, fixed physical embedding, majority-vote chain decoding, and 1000-read budget. A sampled bitstring was counted as a certified candidate only if it represented a feasible vertex cover and its cardinality matched the minimum independently certified by integer linear programming. Across 120 paired comparisons on sparse Barab\'asi--Albert graphs with two attachments per added vertex and graph sizes \(N=70\)--150, DC-OSQA increased the mean number of distinct certified candidates from 171.7 to 455.8 and the mean number of complete-linkage Hamming-distance families from 22.5 to 80.0; paired gains were positive at every tested size. These shifts, together with lower empirical concentration, indicate a change in the composition of the observed certified support. At \(N=150\), the degree-aligned assignment exceeded the within-instance mean over 20 random reassignments of the same offset values in 19 of 20 graphs for certified-support size and all 20 graphs for Hamming families. Within the tested ensemble, these results show that an MVC-specific degree rule changes the certified optima observed within a fixed QPU read budget.
\end{abstract}


\vspace{1em}

\section{Introduction}
Quantum annealing represents an optimization problem as an Ising or quadratic unconstrained binary optimization (QUBO) objective and returns samples from the resulting annealing process \cite{kadowaki1998quantum,farhi2000adiabatic,albash2018adiabatic,lucas2014ising}. When the optimum is degenerate, solver performance is not described fully by the best energy found. The frequencies with which distinct optimal or near-optimal solutions are observed can also matter, particularly when sampled solutions are subsequently screened using criteria that are absent from the original objective \cite{schulhof2025finding,ahanor2024diversitree}.

Uniform sampling over degenerate ground states should not be assumed. Both numerical and hardware studies have shown that transverse-field annealing can strongly favor some optimal configurations over others \cite{matsuda2009groundstate,mandra2017exponentially,konz2019uncertain}. Consequently, the finite-read set of observed optima is a property of the complete sampling procedure, including the encoded Hamiltonian, embedding, hardware calibration, annealing dynamics, and control settings.

Several forms of problem or state information have been used to alter quantum-annealing outcomes. Per-qubit anneal offsets have been tuned iteratively based on qubit floppiness, the empirical probability that flipping a qubit leaves the classical Ising energy unchanged in the sampled configurations, to mitigate perturbative anticrossings and biased sampling of degenerate ground states \cite{lanting2017experimental}. Adame and McMahon instead constructed a static heuristic from the average effective fields implied by the problem Hamiltonian and demonstrated large changes in optimization success probability under inhomogeneous driving \cite{adame2020inhomogeneous}. Inhomogeneous schedules have also been studied algorithmically as controls for solution diversity \cite{mohseni2023sampling}, while bias fields can steer hardware sampling using information about a target state \cite{grass2022sampling}. These studies already establish that problem-informed and state-informed nonuniform controls can reshape annealing outcomes. Here we ask whether a simple rule tied to the constraint incidence of a specific combinatorial problem can change the finite-budget composition of certified optima for an embedded quantum processing unit (QPU) instance, and whether any change remains associated with the intended logical-variable placement under matched controls.

We address this question with Minimum Vertex Cover (MVC). MVC has previously been implemented on quantum-annealing hardware, including through decomposition strategies for larger instances \cite{pelofske2019solving}. Here the objective is different: we study the composition of the finite-read set of optima rather than maximum solvable size. MVC provides a direct link between graph topology and constraint incidence: a vertex of degree \(d_i\) participates in exactly \(d_i\) edge-cover constraints. Vertex-cover algorithms have also been studied on heterogeneous scale-free network models \cite{blasius2023vertex}. Barab\'asi--Albert (BA) graphs provide a controlled sparse ensemble with broad degree heterogeneity \cite{barabasi1999emergence}. This heterogeneity supplies a nontrivial signal for a degree-conditioned control, whereas the rule reduces to a uniform offset on a degree-regular graph. We therefore use the max-normalized degree score \(d_i/d_{\max}\) as a direct measure of local constraint participation. This yields degree-conditioned offset-scheduled quantum annealing (DC-OSQA), in which vertices with larger normalized degree receive more negative anneal offsets.

DC-OSQA tests whether a transparent MVC-specific assignment of local annealing controls changes the composition of the observed certified support within a fixed read budget. This finite-budget objective is related to classical work on recovering multiple exact optima, generating diverse near-optimal solutions, and sampling degenerate ground states more fairly \cite{schulhof2025finding,ahanor2024diversitree,zhu2019fair}. Unlike iterative offset tuning or post hoc diversification, the degree rule is computed once from the graph and evaluated using the same read budget as the zero-offset baseline, QA-Base; it introduces no additional optimization loop within the tested protocol.

Validating membership in the optimum manifold requires more than low QUBO energy or the correct number of selected vertices. For each graph, we independently compute the minimum cover size \(K_{\mathrm{exact}}\) using integer programming. A sampled bitstring is retained as a certified candidate only when its cardinality equals \(K_{\mathrm{exact}}\) and every edge-cover constraint is satisfied. Distinct certified candidates form the certified support, which throughout denotes the set observed at the stated read budget. We additionally summarize their separation in Hamming space using complete-linkage clusters, referred to operationally as Hamming families. Throughout, reshaping certified-optimum sampling refers to changing the finite-read composition of the observed certified optima under a fixed read budget, not to enumerating or uniformly sampling the full optimum manifold.

Our primary comparison uses matched graph instances and deterministic embedding seeds to test DC-OSQA against a zero-offset baseline on sparse Barab\'asi--Albert \(m=2\) graphs. Our contribution is an MVC-specific max-normalized-degree rule for assigning anneal offsets, evaluated through certified-candidate validation and finite-budget support-composition metrics, together with controls that test whether the observed shift remains associated with degree-aligned placement rather than with the offset multiset alone. The primary claim concerns this tested sparse-BA ensemble. Within that scope, the results show that constraint incidence can provide a transparent, precomputed signal for steering which certified optima a quantum annealer returns, without iterative offset tuning.

\section{Methods}
\subsection{Minimum Vertex Cover formulation and certified-candidate validation}

For a graph \(G=(V,E)\) with \(N=|V|\) vertices, the Minimum Vertex Cover (MVC) problem asks for the smallest subset of vertices \(S\subseteq V\) such that every edge \((i,j)\in E\) has at least one endpoint in \(S\). We represent a candidate cover by a binary vector \(x\in\{0,1\}^{|V|}\), where \(x_i=1\) indicates that vertex \(i\) is included in the cover. A feasible vertex cover satisfies
\begin{equation}
x_i+x_j\ge 1,\qquad \forall (i,j)\in E.
\end{equation}

For each graph instance, we computed the minimum vertex-cover size \(K_{\mathrm{exact}}\) using an integer linear programming (ILP) formulation that minimizes \(\sum_{i\in V} x_i\) subject to the edge-cover constraints above and binary constraints \(x_i\in\{0,1\}\). This computation was performed once per graph instance with the general-purpose mixed-integer programming solver HiGHS \cite{huangfu2018parallelizing}, independently of the quantum annealing (QA) and simulated annealing (SA) outputs; it therefore provides a fixed, sampler-independent reference value.

We call a feasible sample whose cover size equals \(K_{\mathrm{exact}}\) a \emph{certified candidate}: a sampled bitstring \(x\) is a certified candidate if it both has cardinality \(K_{\mathrm{exact}}\) and satisfies every edge-cover constraint. Figure~\ref{fig:workflow_dc_osqa}(e) highlights three categories relevant to certification: certified candidates; feasible but nonminimum covers whose cardinality exceeds \(K_{\mathrm{exact}}\); and minimum-cardinality but infeasible bitstrings.  Only certified candidates enter the certified-candidate metrics. The analysis focuses on support discovery within the degenerate optimum manifold. We refer to the set of distinct certified candidates observed at least once as the \emph{observed certified support}. Its size, the unique certified-candidate count, is our primary outcome measure. We distinguish a \emph{certified candidate}, a distinct certified bitstring, from a \emph{certified occurrence}, an individual certified read counted with its multiplicity across the read budget. Support size counts distinct certified candidates, whereas concentration and occurrence-weighted quantities are computed over the empirical distribution of certified occurrences. For a read budget \(R\) containing \(C\) certified occurrences, the certified yield is \(Y=C/R\).

\subsection{QUBO formulation for quantum annealing}

For quantum annealing, MVC was encoded as a quadratic unconstrained binary optimization (QUBO) problem \cite{lucas2014ising}. A vertex cover is represented by a binary vector \(x\in\{0,1\}^{|V|}\), where \(x_i=1\) indicates that vertex \(i\) is selected. We begin with the uncovered-edge penalty form
\begin{equation}
E(x)
=
\lambda_{\mathrm{card}}\sum_{i\in V}x_i
+
\lambda_{\mathrm{edge}}\sum_{(u,v)\in E}(1-x_u)(1-x_v).
\end{equation}
where \(\lambda_{\mathrm{card}}=1.0\) and \(\lambda_{\mathrm{edge}}=2.0\). The first term favors smaller covers, while the second penalizes uncovered edges. With \(\lambda_{\mathrm{edge}}>\lambda_{\mathrm{card}}\), selecting an endpoint of an uncovered edge increases the cardinality term by \(\lambda_{\mathrm{card}}\) but removes at least one penalty \(\lambda_{\mathrm{edge}}\), while creating no new uncovered-edge penalties. An infeasible configuration therefore cannot be a global minimizer.

Expanding the penalty term and omitting the additive constant \(\lambda_{\mathrm{edge}}|E|\) gives the implemented non-constant QUBO
\begin{equation}
E(x)
=
\lambda_{\mathrm{card}}\sum_{i\in V}x_i
-
\lambda_{\mathrm{edge}}\sum_{(u,v)\in E}(x_u+x_v)
+
\lambda_{\mathrm{edge}}\sum_{(u,v)\in E}x_u x_v .
\end{equation}
Collecting its linear terms yields
\begin{equation}
E(x)
=
\sum_{i\in V}
\left(\lambda_{\mathrm{card}}-\lambda_{\mathrm{edge}}d_i\right)x_i
+
\lambda_{\mathrm{edge}}\sum_{(u,v)\in E}x_u x_v ,
\end{equation}
where \(d_i\) is the degree of vertex \(i\). Thus, degree enters the QUBO linear field directly and equals the number of edge-cover constraints incident on the corresponding variable. This dual interpretation motivates the degree-conditioned offset rule used below. All reported certified-candidate analyses nevertheless use the explicit cardinality and feasibility checks defined above rather than QUBO energy alone. Further curation and validation details are provided in the Supplementary Information.

\subsection{Quantum annealing protocols}

The primary comparison is between QA-Base and DC-OSQA. QA-Base is a zero-offset quantum-annealing baseline. DC-OSQA differs only by applying offsets derived from max-normalized vertex degree. Within each graph instance, the two protocols use the same MVC QUBO, fixed physical embedding, read budget, and chain-break decoding procedure. The Supplement reports additional offset controls as secondary comparisons.

On the hardware used here, the anneal is parameterized by a normalized annealing parameter \(s\in[0,1]\) that interpolates between the transverse-field driver and the problem Hamiltonian,
\begin{equation}
H(s)=-\frac{A(s)}{2}\sum_{i}\sigma^{x}_{i}+\frac{B(s)}{2}H_{\mathrm{P}},
\end{equation}
where \(A(s)\) and \(B(s)\) are the device envelope functions and \(H_{\mathrm{P}}\) is the Ising problem Hamiltonian corresponding to the MVC QUBO under the standard binary-to-spin transformation \cite{dwave_annealing_controls}. An anneal offset \(\delta_i\) modifies the control trajectory of qubit \(i\), producing qubit-specific schedules \(A_i(s)\) and \(B_i(s)\) within the device-reported normalized offset range; positive offsets advance the qubit's anneal, whereas negative offsets delay it \cite{dwave_anneal_offsets,lanting2017experimental}.

\begin{figure}[t]
\centering
\includegraphics[width=\textwidth]{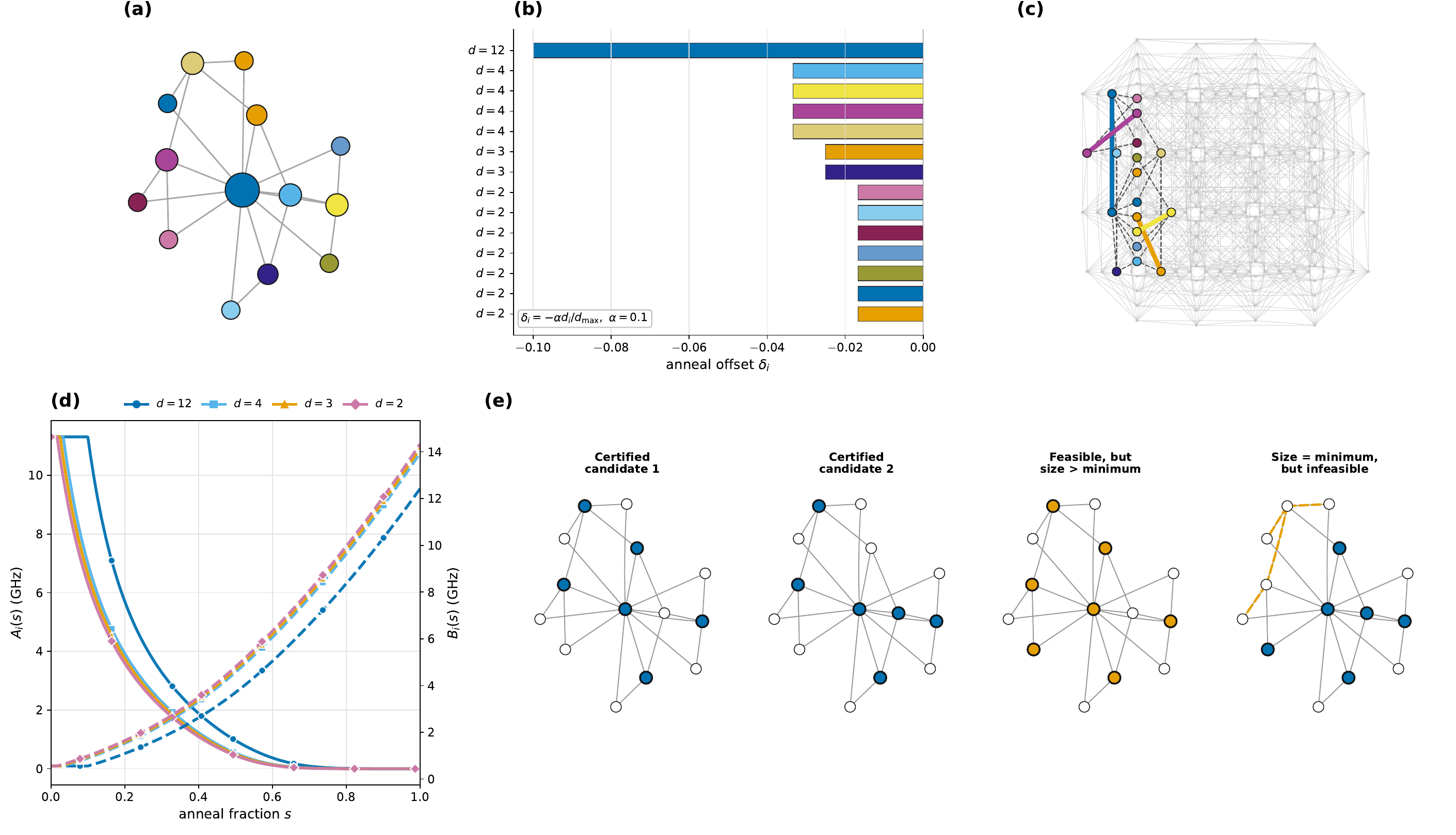}
\caption{
Workflow of degree-conditioned offset-scheduled quantum annealing (DC-OSQA).
(a) Sparse Barab\'asi--Albert \(m=2\) graph with heterogeneous vertex degrees; node size reflects degree, and unique node colors identify logical vertices across panels (a)--(c).
(b) Per-vertex degree-conditioned offset assignment, \(\delta_i=-\alpha d_i/d_{\max}\), with \(\alpha=0.1\); each bar retains the color of its logical vertex.
(c) Minor embedding of the logical graph into a Zephyr hardware graph, with each physical chain shown in the color of its logical vertex.
(d) Offset-dependent schedules for the four degree classes; solid curves show \(A_i(s)\), dashed curves show \(B_i(s)\), and marker shapes distinguish degree classes independently of color.
(e) Four examples spanning the three certification categories: two distinct certified candidates, a feasible but nonminimum cover, and a minimum-cardinality but infeasible bitstring, with uncovered edges in the invalid example marked by orange dashed lines.
}
\label{fig:workflow_dc_osqa}
\end{figure}

Figure~\ref{fig:workflow_dc_osqa} summarizes the DC-OSQA workflow. Starting from an MVC instance on a sparse BA graph, vertex degree is used as a constraint-participation score to assign degree-aligned anneal offsets. These logical offsets are propagated to the physical qubits in each embedded chain, inducing degree-dependent shifts of the nominal annealing schedule. The resulting samples are then filtered using the certified-candidate criterion, which requires both the ILP-certified cardinality \(K_{\mathrm{exact}}\) and explicit vertex-cover feasibility.

\subsection{Quantum annealing hardware, embeddings, and schedules}

Quantum annealing experiments were run on the D-Wave \texttt{Advantage2\_system1} processor \cite{dwave_annealing_controls}, which uses the Zephyr hardware topology \cite{boothby2022zephyr}. For each graph instance, the logical QUBO interaction graph was embedded into the hardware graph using \texttt{minorminer.find\_embedding} \cite{choi2008minor,cai2014practical} with the deterministic instance seed. QA-Base and DC-OSQA shared one minor embedding for each matched graph instance, held fixed using \texttt{Fixed\-Embedding\-Composite}; the two protocols differed only in the applied anneal-offset vector, which was zero for QA-Base and degree-conditioned for DC-OSQA. The retained canonical vertex-to-chain mappings are included in the released data and support the chain-length diagnostic reported in Supplementary Fig.~S5. The same paired-embedding design was used for supplementary centrality-based controls. Vertex-level physical-chain mappings were also retained for the regenerated-embedding sensitivity experiment.

QA-Base used a \(20\,\mu\mathrm{s}\) anneal with no node-dependent offsets, whereas DC-OSQA used the same base anneal with node-dependent offsets derived from the max-normalized degree score. No explicit chain strength was passed in the QPU sampling calls; the Ocean embedding composite's default chain-strength handling was used consistently across matched protocols. Broken chains were decoded by majority vote, and per-sample chain-break fractions, the fraction of embedded chains whose physical qubits disagreed in a given read, were retained for diagnostics.

\subsection{Degree-conditioned anneal offsets}

For each graph, we computed a max-normalized degree score \(c_i\in[0,1]\) for each logical variable:
\begin{equation}
c_i = \frac{d_i}{d_{\max}},
\qquad d_{\max}=\max_j d_j,
\end{equation}
where \(d_i\) is the degree of vertex \(i\). We use ``degree score'' rather than the standard graph-theoretic degree-centrality normalization \(d_i/(N-1)\), because the implemented rule is normalized by the instance maximum.

Each logical variable received an anneal offset \(\delta_i=-\alpha c_i\), with \(\alpha=0.1\) in the main experiments. We treated \(\alpha=0.1\) as the main operating point and evaluated smaller and larger offset scales in the Supplement.

For MVC, \(d_i\) counts the edge-cover constraints involving \(x_i\) and also enters its collected QUBO linear coefficient. DC-OSQA therefore defines the monotone map \(d_i>d_j\Rightarrow\delta_i<\delta_j\). Under the hardware convention used here, the more negative offsets assigned to higher-degree variables delay the corresponding qubits relative to the nominal annealing trajectory.

DC-OSQA tests whether differentiating local annealing trajectories according to constraint participation alters the relative sampling frequencies of certified optima and, potentially, the support observed within a fixed read budget.

The same logical offset was assigned to all physical qubits in the embedded chain for that logical variable. Requested physical-qubit offsets were checked against the QPU-reported \texttt{anneal\_offset\_ranges} \cite{dwave_anneal_offsets}; offset settings outside the allowed hardware range were not used. Additional offset-scale checks were retained as sensitivity analyses in the Supplement.

\subsection{Primary ensemble and canonical acquisition}
\label{sec:primary_ensemble}

The Barab\'asi--Albert (BA) model adds each new vertex with \(m\) edges to existing vertices selected preferentially by degree \cite{barabasi1999emergence}; thus, \(m\) controls the minimum degree and the asymptotic mean degree \(2m\). The main experiments used sparse BA graphs with \(m=2\). Graphs were generated with NetworkX using the deterministic seed \(42+1000N+i\), where \(i\) is the zero-based instance index; the resulting graph and embedding seeds were retained in the released metadata. The matched QA-Base and DC-OSQA conditions therefore share both graph identity and the recorded physical embedding. The canonical dataset includes
\[
N\in\{20,30,40,50,60,70,80,90,100,120,150\}.
\]

We used 10 graph instances at each size for \(N=20\)--60 and 20 instances at each size for \(N=70\)--150. Each canonical QPU protocol--instance pair used 1000 reads.

Sparse BA graphs provide a tractable testbed with a broad relative degree range, practical embeddability, and multiple observed minimum covers. The wide range of \(d_i/d_{\max}\) also allows the offset rule to differentiate variables at fixed \(\alpha\). This choice defines the scope of the study; denser BA ensembles and a degree-regular negative control are treated as scope tests.

For every degree-group summary on the BA \(m=2\) ensemble, we used the fixed actual-degree ranges low (\(d\leq2\)), middle (\(3\leq d\leq4\)), and high (\(d\geq5\)). In this model, degree two is the minimum and modal class, and the mean degree approaches \(2m=4\). The ranges therefore separate the modal minimum-degree class, degrees three through the asymptotic mean, and the above-mean tail. This ensemble-based definition was applied unchanged across \(N\), rather than recomputed from graph-specific quantiles. Across the 20 graphs at each \(N=70\)--150, the three groups contained 47.4--50.0\%, 28.5--31.1\%, and 21.1--22.4\% of vertices, respectively. The same definition was retained for the \(N=150\) matched-shuffle diagnostic so that its grouped summary is directly comparable with the canonical analysis. These labels are specific to BA \(m=2\) and were not transferred to the denser BA, ER, or degree-regular controls.

\subsection{Extension and control experiments}

Table~\ref{tab:experiment_map} separates the primary comparison from the extensions, controls, sensitivities, and complementary diagnostics. QPU budgets are reads per protocol--instance condition, whereas SA budgets are output samples.

\begin{table}[htbp]
\centering
\footnotesize
\setlength{\tabcolsep}{3.2pt}
\renewcommand{\arraystretch}{1.10}
\caption{Map of the primary experiment, extensions, controls and sensitivity analyses, and diagnostics.}
\label{tab:experiment_map}
\begin{tabular}{>{\raggedright\arraybackslash}p{2.95cm}>{\raggedright\arraybackslash}p{4.25cm}>{\raggedright\arraybackslash}p{4.95cm}>{\raggedright\arraybackslash}p{2.55cm}}
\toprule
Component & Graphs and units & Comparison or varied factor & Acquisition or calculation \\
\midrule
\multicolumn{4}{@{}l}{\itshape Primary experiment}\\
\addlinespace[1pt]
Canonical comparison & BA \(m=2\), \(N=20\)--150;\newline 10 graphs per size for \(N=20\)--60 and 20 for \(N=70\)--150 & QA-Base vs DC-OSQA at \(\alpha=0.1\) & QPU, 1000 reads \\
\midrule
\multicolumn{4}{@{}l}{\itshape Extensions}\\
\addlinespace[1pt]
Graph size & BA \(m=2\), \(N=200\);\newline 20 graphs & Canonical QA-Base--DC-OSQA comparison at larger \(N\) & QPU, 1000 reads \\
\addlinespace[3pt]
Read budget & BA \(m=2\), \(N=150\);\newline 20 graphs & First \(R\) reads from each 50,000-read stream, with \(R=1000\), 2000, 5000, 10,000, 20,000, and 50,000 & QPU, 50,000 total reads \\
\midrule
\multicolumn{4}{@{}l}{\itshape Controls and sensitivities}\\
\addlinespace[1pt]
Score and offset scale & BA \(m=2\), \(N=70\)--150 & Degree vs betweenness score; \(\alpha\in\{0.05,0.1,0.2\}\) & QPU, 1000 reads \\
\addlinespace[3pt]
Offset assignment & BA \(m=2\), \(N=150\);\newline 20 graphs & Degree-aligned assignment vs 20 random reassignments of the same offset multiset & QPU, 1000 reads \\
\addlinespace[3pt]
Graph density & BA \(m=3,4\), \(N=70,90,120,150\); 10 graphs per size & \(m\) and \(\alpha\in\{0.01,0.025,0.05,0.10\}\) varied & QPU, 1000 reads \\
\addlinespace[3pt]
Graph topology & Random 3-regular, \(N=150\), 20 graphs; connected Erd\H{o}s--R\'enyi (ER), \(N=150\), 10 graphs & Degree-regular boundary and connected-ER topology transfer & QPU, 1000 reads \\
\midrule
\multicolumn{4}{@{}l}{\itshape Diagnostics}\\
\addlinespace[1pt]
Embedding & BA \(m=2\), \(N=150\);\newline 20 graphs & Five regenerated embeddings per graph; QA-Base and DC-OSQA paired on each embedding & QPU, 1000 reads \\
\addlinespace[3pt]
Classical recovery & BA \(m=2\), \(N=150\);\newline 20 graphs & SA recovery and QPU-support overlap at matched output budgets & SA, 1,000, 10,000, and 50,000 samples \\
\addlinespace[4.5pt]
Idealized closed-system model & BA \(m=2\), \(N=12\)--20;\newline 32--48 graphs per size with at least four exact minimum covers & Degree-aligned vs zero-offset and assignment controls & Full-state-vector evolution, dimensionless \(T=40\) \\
\bottomrule
\end{tabular}
\end{table}

\paragraph{Size and read-budget extensions.}
The \(N=200\) experiment used 20 BA \(m=2\) graph instances and was analyzed separately from the canonical dataset and the primary pooled \(N=70\)--150 summary. For the \(N=150\) read-budget extension, each QA-Base and DC-OSQA protocol--instance pair used one 50,000-read stream. At each evaluated budget \(R\in\{1000,2000,5000,10000,20000,50000\}\), certified-support size and Hamming-family count were evaluated using the first \(R\) reads in acquisition order; means and standard errors were computed across the 20 graph instances.

\paragraph{Score and offset-scale controls.}
Betweenness-centrality OSQA (BC-OSQA) replaced the degree score with maximum-normalized betweenness centrality on the canonical BA \(m=2\), \(N=70\)--150 subset. Offset-scale sensitivity was evaluated at \(\alpha\in\{0.05,0.1,0.2\}\) on matched BA \(m=2\) instances at \(N\in\{70,90,120,150\}\). These controls test dependence on the structural score and offset scale, respectively.

\paragraph{Offset-assignment control.}
For each of 20 BA \(m=2\), \(N=150\) graphs, the degree-aligned assignment was compared with 20 random reassignments of the same logical offset multiset on a fixed fresh embedding. Each reassignment was retained only when all resulting physical-qubit offsets lay within the device-reported allowed ranges. This control tests dependence on logical-variable placement.

\paragraph{Embedding diagnostic.}
Five new minor embeddings were generated for each of the 20 canonical BA \(m=2\), \(N=150\) graphs, giving 100 graph--embedding pairs. QA-Base and DC-OSQA were paired on each regenerated embedding, and the corresponding vertex-level mappings were retained.

\paragraph{Density and topology controls.}
Denser BA controls used \(m\in\{3,4\}\), \(N\in\{70,90,120,150\}\), 10 graphs per size, and \(\alpha\in\{0.01,0.025,0.05,0.10\}\). The degree-regular boundary control used 20 random 3-regular graphs at \(N=150\); because every vertex has degree three, DC-OSQA reduces to the uniform offset \(\delta_i=-0.1\). The topology-transfer control used 10 connected Erd\H{o}s--R\'enyi graphs at \(N=150\), generated with \(p=4/(N-1)=4/149\), which targets an unconditional mean degree of four. Disconnected ER draws were rejected using deterministic seed increments, and accepted seeds were retained. The ER comparison therefore targets nominal mean degree; realized edge counts and reconstruction details are reported in the Supplement. All three controls used matched QA-Base and DC-OSQA conditions with 1000 reads per protocol.

\paragraph{Classical recovery probe.}
Simulated annealing was evaluated on the same 20 BA \(m=2\), \(N=150\) graphs as a certified-support recovery and overlap probe. Its implementation and output budgets are specified in Sec.~\ref{sec:sa_recovery}.

\paragraph{Idealized closed-system check.}
A full-state-vector calculation on BA \(m=2\) surrogates with \(N=12\)--20 tested whether degree-aligned local schedules alone can redistribute occupation within a degenerate MVC optimum manifold, using a zero-offset baseline and four assignment controls. Full model and analysis details are given in Supplementary Sec.~S8.

\subsection{Study design and outcome hierarchy}

The primary estimand is the graph-instance-paired difference between DC-OSQA and QA-Base in the number of unique certified candidates observed within the canonical 1000-read budget. The main larger-size summary is reported for the sparse BA \(m=2\) subset with \(N=70\)--150, with size-specific results retained to avoid relying only on the pooled comparison. Complete-linkage Hamming-family count is a secondary, threshold-dependent structural summary of the observed candidate set. Concentration metrics and degree-resolved inclusion shifts are interpretive diagnostics: they characterize the composition of certified occurrences and are evaluated only where their conditional definitions are meaningful.

The canonical comparison in Table~\ref{tab:experiment_map} is primary; all other components are extensions, sensitivities, controls, or complementary diagnostics. Size and budget extensions test persistence beyond the canonical acquisition conditions, the controls probe robustness and transfer, SA probes set recovery, and the idealized calculation probes model-level sufficiency. Unless otherwise stated, hardware diagnostic \(p\)-values are unadjusted for multiplicity and are interpreted descriptively with effect direction, magnitude, and uncertainty.

\subsection{Data curation and zero-yield retention}

Raw QPU and simulated-annealing outputs were curated into analysis-ready tables used for the reported manuscript statistics. For each protocol--instance pair, we report the unique certified-candidate count. Protocol--instance pairs with no certified candidates were retained as zero-yield observations rather than dropped, preventing performance inflation from conditioning the analysis on successful runs. Additional curation and audit details are reported in the Supplement.

\subsection{Observed-support diversity and concentration metrics}
\label{sec:diversity_concentration_metrics}

We summarized the certified support using two metrics. First, we counted the unique certified candidates observed for each protocol on each graph instance. Second, we grouped these candidates into Hamming families.

For two bitstrings \(x,y\in\{0,1\}^N\), the Hamming distance is
\begin{equation}
d_H(x,y)=\sum_{i=1}^{N}|x_i-y_i|.
\end{equation}

Hamming families were computed by complete-linkage hierarchical clustering \cite{murtagh2012algorithms} over unique certified candidates. The reported metric uses the fixed operational threshold \(\tau=0.1N\).

Complete linkage defines the distance between two clusters \(C_a\) and \(C_b\) as
\begin{equation}
D(C_a,C_b)=\max_{x\in C_a,\,y\in C_b} d_H(x,y).
\end{equation}

Thus, cutting the complete-linkage dendrogram at threshold \(\tau\) ensures that any two candidates within the same family differ by at most \(\tau\) bits. This makes the resulting clusters a conservative operational summary of certified-candidate diversity.

 Complete linkage was adopted because a threshold cut directly enforces the intended mutually-nearby interpretation. These families remain threshold-dependent Hamming-space summaries, not physical energy basins or dynamical states.

For concentration summaries, index the \(M\) distinct observed certified candidates by \(j=1,\ldots,M\). Let \(n_j\) be the occurrence count of candidate \(j\), \(C=\sum_{j=1}^{M}n_j\) the total certified occurrences, and \(p_j=n_j/C\) its empirical occurrence probability, so that \(\sum_jp_j=1\). We report normalized entropy
\begin{equation}
H_{\mathrm{norm}}=
\begin{cases}
-\displaystyle\frac{\sum_j p_j\log p_j}{\log M}, & M>1,\\[4pt]
0, & M=1,
\end{cases}
\end{equation}
the Simpson effective-support fraction \(1/(M\sum_jp_j^2)\), the top-decile share obtained by summing \(p_j\) over the \(\lceil0.1M\rceil\) most frequent candidates, and the Gini coefficient
\begin{equation}
G=\frac{\sum_{j,k}|p_j-p_k|}{2M}.
\end{equation}
These metrics are conditional on observing at least one certified occurrence; degenerate-case conventions and additional concentration summaries are given in the Supplement.

\subsection{Simulated annealing recovery and overlap diagnostics}
\label{sec:sa_recovery}

Simulated annealing \cite{kirkpatrick1983optimization} served as a finite-budget recovery and overlap diagnostic. The analysis used the same 20 BA \(m=2\), \(N=150\) graph instances, independently certified \(K_{\mathrm{exact}}\) values, and explicit feasibility filter as the QPU analysis. Three SA acquisitions entered the overlap analyses: a separately retained 1,000-sample SA-Max dataset and two independent recovery datasets with output budgets of 10,000 and 50,000 samples. All used 10,000 sweeps and an inverse-temperature range \(\beta\in[0.1,5.0]\). For overlap analyses, we compare observed certified-candidate sets using recovery fractions, Jaccard overlap, sampler-specific candidates, and union gain over the better individual sampler. These quantities summarize finite-budget recovery and partial overlap between observed certified-support sets; detailed set definitions are provided in the Supplement. Because the SA and QPU acquisitions were not runtime-normalized, this comparison evaluates certified-support recovery and overlap between finite sample sets rather than ranking computational performance.

\subsection{Statistical analysis}

All protocol comparisons were paired by graph instance. The descriptive larger-size subset \(N=70\)--150 comprises 120 paired comparisons across six graph sizes, and results are also reported by size. Wins, losses, and ties were counted at the graph-instance level. Two-sided Wilcoxon signed-rank tests were applied to paired differences, with zero differences omitted from the signed-rank statistic but retained as ties in the descriptive counts.

For the primary outcomes, paired standardized effects were calculated as
\begin{equation}
d_z=\frac{\overline{\Delta}}{s_{\Delta}},
\end{equation}
where \(\Delta\) is the graph-level DC-OSQA minus QA-Base difference and \(s_{\Delta}\) is its sample standard deviation. Percentile-bootstrap 95\% confidence intervals for \(\overline{\Delta}\) and \(d_z\) were obtained from 5,000 graph-size-stratified resamples. Within each \(N\) stratum, graph-instance pairs were sampled with replacement while retaining the original number of instances; the 2.5th and 97.5th percentiles of the resulting statistic defined the interval. Exploratory diagnostics and their comparison units are described with the corresponding supplementary analyses.

\subsection{Reproducibility}

Primary statistics were computed from the canonical curated dataset, and supplementary diagnostics were computed from their corresponding released analysis inputs. The Supplementary Information documents zero-yield retention, expanded implementation details, and robustness analyses; the data release provides the associated analysis inputs and figure-generation scripts.

\section{Results}
\subsection{Certified-support expansion under degree-conditioned offsets}

\begin{figure}[htbp]
\centering
\includegraphics[width=0.88\textwidth,keepaspectratio]{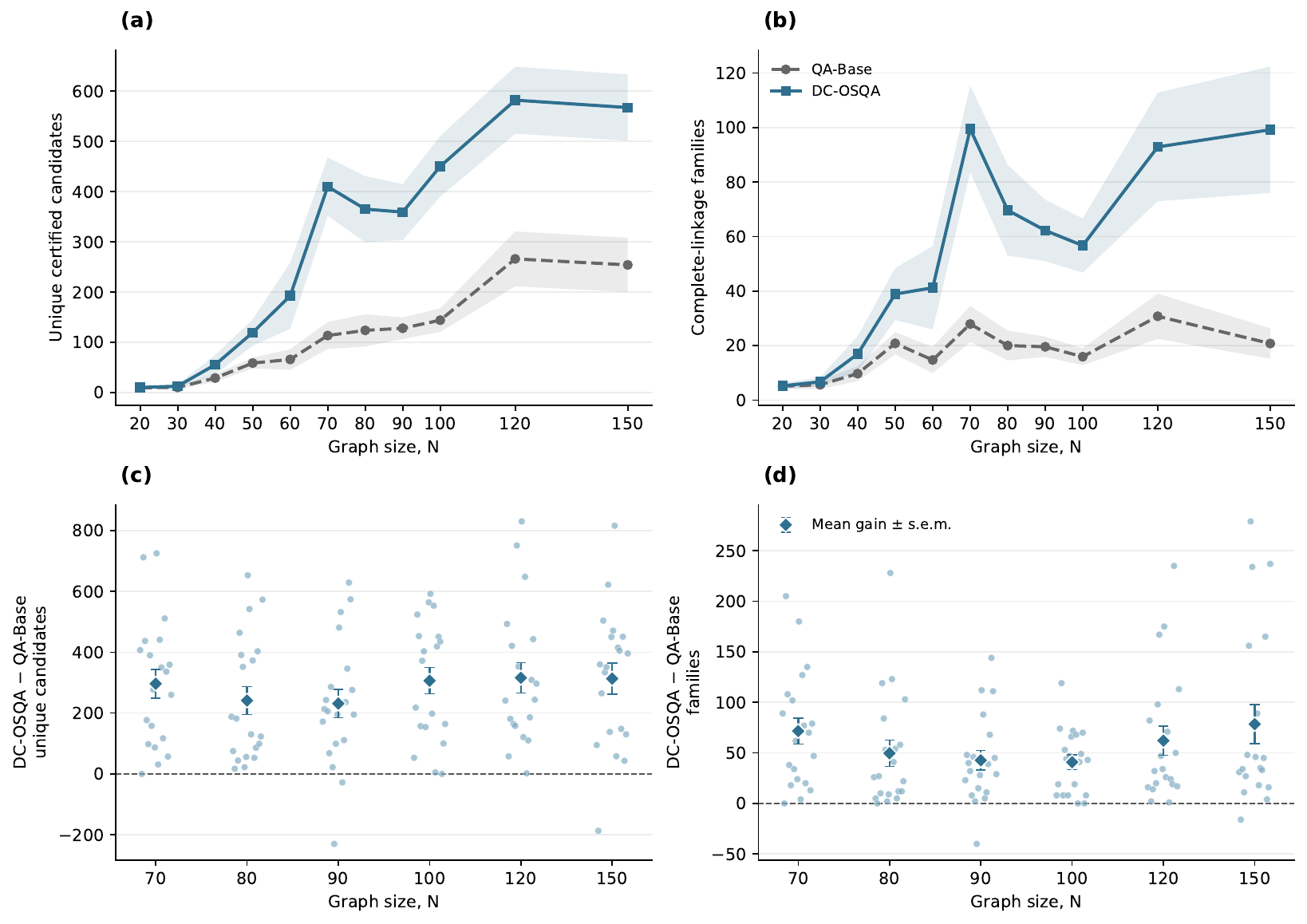}
\caption{
Degree-conditioned anneal offsets broaden the certified support in sparse Barab\'asi--Albert MVC instances.
(a) Mean number of unique certified candidates discovered by QA-Base and DC-OSQA.
(b) Mean number of complete-linkage Hamming families with threshold \(\tau=0.1N\).
(c,d) Instance-paired DC-OSQA minus QA-Base gains for the same two metrics in the larger-size subset, \(N=70\)--150.
In (a,b), shaded bands show \(\pm\) standard error of the mean (s.e.m.) across graph instances. In (c,d), small blue circles show individual paired gains and diamonds with error bars show the mean \(\pm\) s.e.m. at each graph size.
All counts use certified-candidate validation: candidates must both match \(K_{\mathrm{exact}}\) and satisfy vertex-cover feasibility.
}
\label{fig:main_qa_dc}
\vspace{-0.4\baselineskip}
\end{figure}

We first test whether a degree-conditioned offset rule changes the certified support within a fixed read budget. Sparse Barab\'asi--Albert \(m=2\) MVC instances combine multiple observed optima with a broad degree range, allowing max-normalized degree to serve as a constraint-participation signal. Figure~\ref{fig:main_qa_dc} summarizes the primary comparison between QA-Base and DC-OSQA on this ensemble.

DC-OSQA increased the number of unique certified candidates relative to QA-Base, with positive mean paired gains at every tested graph size from \(N=70\) to \(N=150\) (Fig.~\ref{fig:main_qa_dc}). Mean paired gains were also positive at every smaller size, growing from 0.6 at \(N=20\) to 126.7 at \(N=60\), with a single negative paired difference among the 50 smaller-size pairs. The \(N=70\)--150 subset therefore restricts the pooled summary to sizes with larger effect magnitudes rather than marking a change in direction. Protocol--instance pairs with no certified candidates were retained as zeros, following the zero-yield convention described in the Methods.

We next asked whether the support expansion also spanned more separated regions of Hamming space. DC-OSQA increased the complete-linkage Hamming-family count, with positive mean paired gains at all six larger graph sizes (Fig.~\ref{fig:main_qa_dc}); the observed difference is therefore visible in both certified-support size and the secondary Hamming-space summary.

The paired nature of the improvement is shown in Fig.~\ref{fig:main_qa_dc}(c,d). Most matched instances have positive differences for both unique certified candidates and complete-linkage Hamming families. Mean effects vary nonmonotonically with \(N\), and the few non-positive gains occur mainly in lower-yield settings. We therefore interpret the result as an ensemble-level paired effect. Table~\ref{tab:paired_stats} consolidates the protocol means, paired differences, standardized effects, and paired tests for the larger-size subset.
\begin{table}[htbp]
\centering
{
\caption{
Instance-paired statistics for the main sparse BA \(m=2\) setting, pooled over \(N=70\)--150. Differences and wins/losses/ties (W/L/T) are for DC-OSQA relative to QA-Base. Confidence intervals are graph-size-stratified bootstrap intervals; \(p\)-values are from two-sided paired Wilcoxon tests.
}
\label{tab:paired_stats}
\small
\resizebox{\textwidth}{!}{%
\begin{tabular}{lrrrrrr}
\toprule
Metric & \shortstack{QA-Base\\mean} & \shortstack{DC-OSQA\\mean} & \shortstack{Mean paired gain\\(95\% CI)} & \shortstack{\(d_z\)\\(95\% CI)} & W/L/T & \shortstack{Wilcoxon\\\(p\)} \\
\midrule
Unique certified candidates & 171.7 & 455.8 & \shortstack{284.08\\(246.58--320.67)} & \shortstack{1.34\\(1.16--1.57)} & 115/3/2 & \(5.80\times10^{-20}\) \\
Complete-linkage families & 22.5 & 80.0 & \shortstack{57.56\\(47.39--68.25)} & \shortstack{0.96\\(0.85--1.10)} & 114/2/4 & \(7.12\times10^{-20}\) \\
\bottomrule
\end{tabular}
}
}
\end{table}

Both outcomes favored DC-OSQA with positive paired differences and standardized effects whose 95\% confidence intervals excluded zero. The pooled paired Wilcoxon tests gave \(p=5.80\times10^{-20}\) for unique certified candidates and \(p=7.12\times10^{-20}\) for Hamming families (Table~\ref{tab:paired_stats}).

\FloatBarrier
\subsection{Observed certified-sample concentration}

\begin{figure}[htbp]
\centering
\includegraphics[width=0.84\textwidth,keepaspectratio]{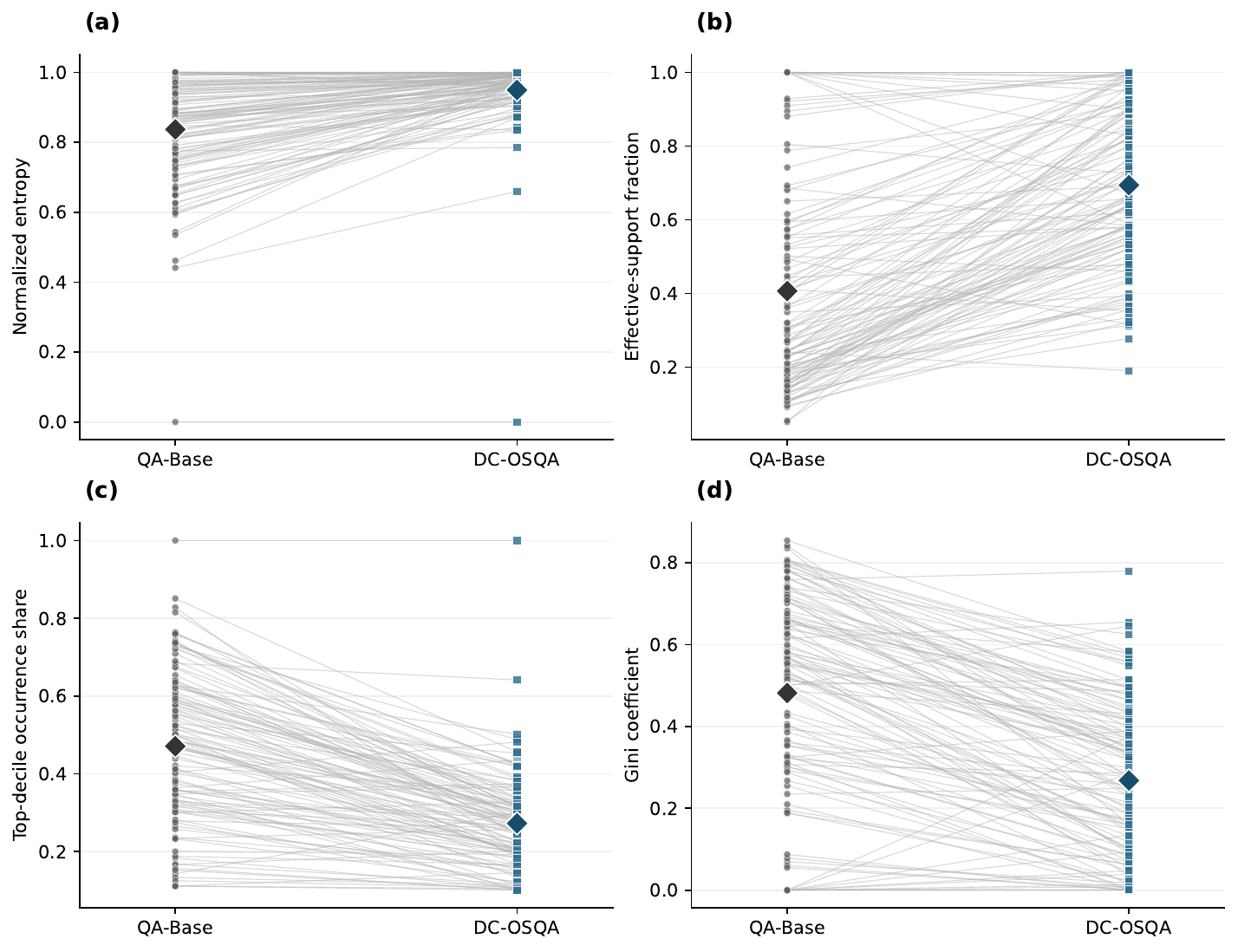}
\caption{
Observed certified-sample concentration diagnostics for BA \(m=2\), \(N=70\)--150.
Each line connects a matched graph instance for QA-Base and DC-OSQA.
The panels show normalized entropy, Simpson effective-support fraction, top-decile occurrence share, and the Gini coefficient over the observed certified occurrences.
Large diamonds mark protocol means across the 118 matched instances. Higher entropy and effective-support fraction, together with lower top-decile share and Gini coefficient, indicate lower concentration in the observed certified occurrences.
}
\label{fig:concentration}
\end{figure}

Figure~\ref{fig:concentration} examines whether the increase in unique certified candidates was accompanied by a change in concentration within the observed certified occurrences. The four metrics were computed from the empirical certified-occurrence distribution using the definitions in Sec.~\ref{sec:diversity_concentration_metrics} and were evaluated only for protocol--instance pairs with at least one certified occurrence.

Across \(N=70\)--150, DC-OSQA had higher normalized entropy than QA-Base in 106 of 118 paired comparisons and higher Simpson effective-support fraction in 100 of 118 comparisons. Conversely, DC-OSQA had lower top-decile share in 108 of 118 comparisons and lower Gini coefficient in 100 of 118 comparisons. The mean normalized entropy increased from 0.84 for QA-Base to 0.95 for DC-OSQA, while the mean top-decile share decreased from 0.47 to 0.27. All four metrics shifted in the lower-concentration direction, with paired Wilcoxon \(p\leq3.77\times10^{-17}\) (Supplementary Table~S1). Conditional on observing certified candidates, these results indicate lower empirical concentration on dominant observed candidates.

\FloatBarrier
\subsection{Degree-aligned composition and placement controls}

To relate the composition change to the graph feature used for offset assignment, we compared vertex-inclusion probabilities using the fixed degree groups defined in Sec.~\ref{sec:primary_ensemble}. The comparison used both equal weighting of distinct certified candidates and occurrence weighting. Inclusion decreased for \(d\leq2\) and increased for both higher-degree groups, with the largest mean increase at \(d=3\)--4 (Table~\ref{tab:degree_shifts}; Supplementary Fig.~S8).

\begin{table}[htbp]
\centering
{
\caption{Degree-resolved changes in certified-candidate inclusion for sparse BA \(m=2\), \(N=70\)--150. Entries are mean DC-OSQA minus QA-Base inclusion-probability shifts across the 118 matched graph-instance comparisons for which both conditional summaries were defined. Supplementary Fig.~S8 shows the instance-level distributions and s.e.m. error bars.}
\label{tab:degree_shifts}
\small
\begin{tabular}{lrr}
\toprule
Actual-degree group & Distinct-candidate weighting & Occurrence weighting \\
\midrule
Low (\(d\leq2\)) & \(-0.04\) & \(-0.05\) \\
Middle (\(3\leq d\leq4\)) & \(+0.05\) & \(+0.06\) \\
High (\(d\geq5\)) & \(+0.02\) & \(+0.03\) \\
\bottomrule
\end{tabular}
}
\end{table}
We also examined vertex-level chain length on the canonical \(N=70\)--150 mappings. Degree and chain length were positively correlated in all 120 graph instances (mean Spearman \(\rho=0.67\)). Among the 118 instances for which both protocols produced certified occurrences, the unadjusted degree--inclusion-shift association was positive in every instance (mean \(\rho=0.23\)). After rank-residualizing chain length, this association remained positive in 117 of 118 instances (mean partial \(\rho=0.24\)); the complementary chain-length association after rank-residualizing degree averaged \(-0.11\). This descriptive partial-rank analysis addresses additive monotone chain-length association. Full definitions and distributions are reported in Supplementary Fig.~S5.

\begin{figure}[H]
\centering
\includegraphics[width=0.93\textwidth,keepaspectratio]{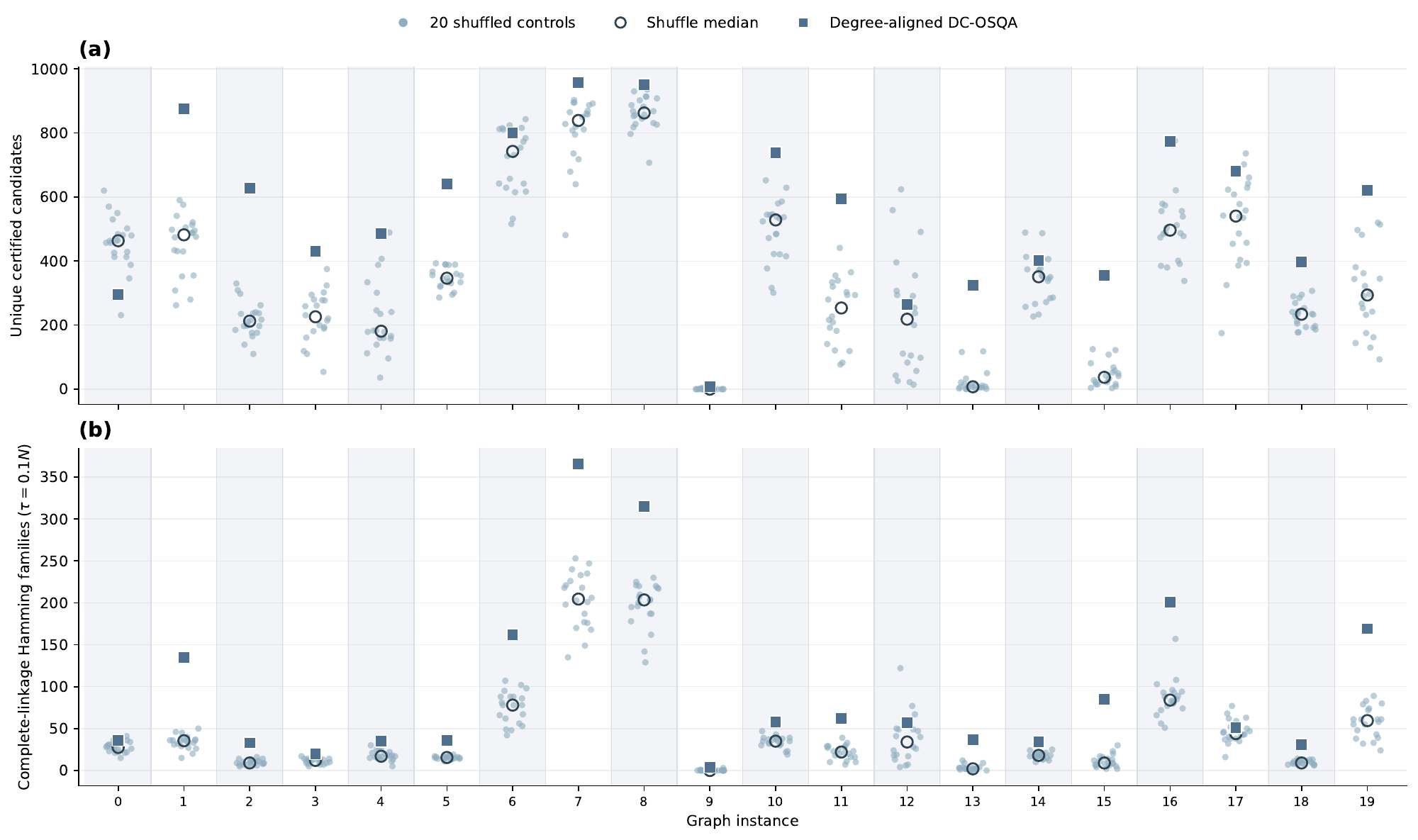}
\caption{
Degree-aligned offset placement compared with random reassignments of the same offset multiset at \(N=150\).
For each of 20 BA \(m=2\) graph instances, light points show 20 hardware-valid shuffled assignments, open circles show the within-instance shuffle median, and dark squares show the original degree-aligned DC-OSQA assignment.
The graph, QUBO, read budget, embedding, offset-value multiset, and certified-candidate rule are held fixed within each instance.
(a) Unique certified candidates.
(b) Complete-linkage Hamming families at threshold \(\tau=0.1N\).
The degree-aligned assignment generally lies above the within-instance shuffle median.
}
\label{fig:topology_assignment_control}
\end{figure}

We next tested whether the degree-to-variable assignment contributed beyond the presence of the same inhomogeneous offset multiset. In a separate \(N=150\) control, each graph was evaluated with the original degree-aligned assignment and 20 hardware-valid random reassignments of the same logical offset values, using a fresh embedding held fixed within that graph. Relative to the within-instance mean over shuffled assignments, degree-aligned DC-OSQA increased the unique certified-candidate count in 19 of 20 graphs, with a mean gain of 198.76 candidates and a two-sided Wilcoxon \(p=4.8\times10^{-5}\). It increased the complete-linkage Hamming-family count in all 20 graphs, with a mean gain of 50.44 families and \(p=2.0\times10^{-6}\). Against the best of the 20 shuffles, the unique-candidate difference was not significant, whereas the Hamming-family difference remained positive in 17 of 20 graphs. Thus, this control supports degree-aligned placement relative to the sampled reassignment distribution. Figure~\ref{fig:topology_assignment_control} shows the instance-level distributions; full shuffle summaries and degree--inclusion correlations are reported in Supplementary Table~S4 and Fig.~S6.

\subsection{QPU read-budget extension at \texorpdfstring{\(N=150\)}{N=150}}
\label{sec:finite_read_support_growth}

To test whether the DC-OSQA effect was restricted to the 1000-read regime, we performed an additional QPU read-budget extension at \(N=150\) on the same 20 BA \(m=2\) instances, separate from the canonical 1000-read runs reported above. Each protocol--instance pair was extended to 50,000 reads, using the same graph instances, protocol definitions, and cached embeddings.

\begin{figure}[tbp]
\centering
\begin{minipage}{0.94\textwidth}
\centering
\includegraphics[width=\linewidth,keepaspectratio]{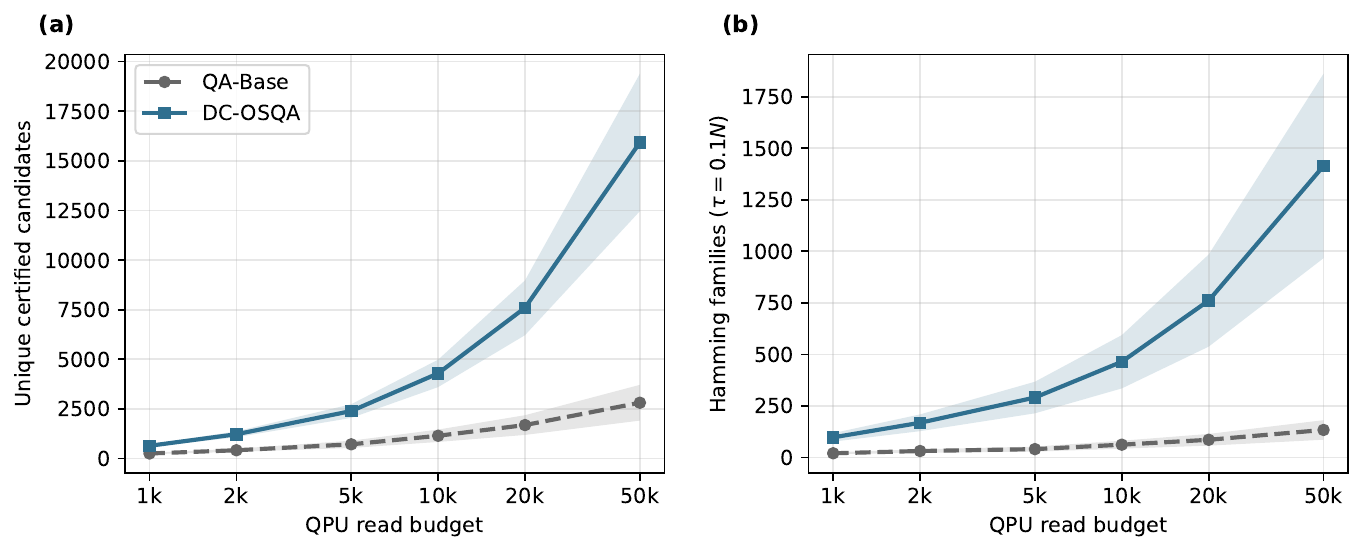}
\caption{
QPU read-budget extension at \(N=150\) on the same 20 BA \(m=2\) graph instances. (a) Unique certified candidates. (b) Complete-linkage Hamming families at threshold \(\tau=0.1N\). Each point summarizes the cumulative acquisition-order prefix at the stated read budget. Lines show across-instance means, and shaded regions show the s.e.m.
QA-Base is shown by gray dashed lines with circles, and DC-OSQA by blue solid lines with squares.
}
\label{fig:qpu_budget_growth}
\end{minipage}
\end{figure}

At 50,000 reads, DC-OSQA discovered more unique certified candidates than QA-Base on 19 of 20 matched instances. The mean number of unique certified candidates was \(15{,}929.55\) for DC-OSQA and \(2{,}812.20\) for QA-Base, corresponding to a mean gain of \(13{,}117.35\) candidates and a paired Wilcoxon \(p=3.81\times10^{-6}\) (Supplementary Table~S6). Across the evaluated budgets from 1000 to 50,000 reads, DC-OSQA also maintained a larger mean certified-support size and more complete-linkage Hamming families than QA-Base (Fig.~\ref{fig:qpu_budget_growth}).

The 50,000-read QPU support-overlap analysis is reported in Supplementary Table~S7; the retained 10,000-read chain-break diagnostic is reported in Supplementary Table~S8.

\FloatBarrier
\subsection{Robustness, transfer boundaries, and complementary checks}

The controls distinguish robustness within the sparse BA setting from transfer beyond it: positive gains persisted across alternative structural scores and regenerated embeddings, their magnitude depended on offset scale, and the sparse-BA direction reversed in the 3-regular and connected-ER ensembles. BC-OSQA produced positive mean gains over QA-Base, but DC-OSQA produced larger gains at every tested size for both primary outcomes (Supplementary Fig.~S2). Both outcomes varied with the offset scale \(\alpha\) (Supplementary Fig.~S3). Across five regenerated embeddings for each of 20 \(N=150\) graphs, graph-level DC-OSQA gains were positive for all 20 graphs in both outcomes (Supplementary Fig.~S4). In denser BA ensembles, the preferred scale shifted: BA \(m=3\) often favored \(\alpha=0.05\) at \(N\geq90\), whereas \(\alpha=0.1\) became detrimental for BA \(m=4\) at \(N=120\) and 150 (Supplementary Fig.~S7).

The topology controls did not reproduce the sparse-BA response. On 20 random 3-regular graphs, QA-Base exceeded DC-OSQA in 19 instances for certified-support size and in 18 for Hamming families, with one family-count tie (Supplementary Table~S5 and Fig.~S9). On 10 connected ER graphs with nominal mean degree four, mean certified-support size was 19.9 for QA-Base and 3.5 for DC-OSQA (mean paired gain \(-16.4\), \(p=2.34\times10^{-2}\)). The ER comparison had low certified yield under both protocols and is treated as a small-sample transfer boundary (Supplementary Fig.~S10). At \(N=200\), DC-OSQA retained positive paired gains for both outcomes, while BC-OSQA showed a positive family-count gain but a less consistent certified-support-size gain (Supplementary Table~S2).

Separately, simulated annealing served as a supplementary recovery and set-overlap probe. At a matched output budget of 50,000 samples, SA yielded a larger mean certified-support size than DC-OSQA (19,040.15 versus 15,929.55), while their supports overlapped only partially (Supplementary Tables~S3 and S7; Fig.~S1). The SA analysis thus provides classical context for the offset-dependent change in QPU sampling composition rather than a computational-performance comparison.

Across all five sizes in the idealized closed-system calculation, the degree-aligned schedule broadened conditional target-manifold support relative to the zero-offset baseline and shuffled assignments and produced positive intermediate-budget expected-discovery contrasts (Supplementary Figs.~S12 and S13). Curve-wise tests survived correction at four sizes; the \(N=18\) contrast remained positive descriptively but was not significant after correction. This establishes sufficiency within the specified model rather than reproducing the QPU dynamics.

Together, these checks delimit the empirical scope of the observed effect: the sparse-BA response was robust to several within-ensemble perturbations but depended on offset scale, graph density, and topology.

\FloatBarrier

\section{Discussion}

The central result is that a constraint-derived local control changes the finite-read certified support observed from a quantum annealer. Within the tested sparse BA \(m=2\) ensemble, it changes which certified optima are observed and how concentrated their empirical occurrence distribution is. This distinction is important whenever downstream use depends on receiving multiple alternatives rather than repeated copies of one dominant optimum.

\subsection{Degree-aligned redistribution}

The observed composition shift is directionally consistent with the offset rule. DC-OSQA reduced the inclusion probability of vertices with \(d\leq2\) and increased it in both the \(d=3\)--4 and \(d\geq5\) groups under both weighting schemes (Table~\ref{tab:degree_shifts}; Supplementary Fig.~S8). The largest mean increase occurred at \(d=3\)--4, while both groups with \(d\geq3\) shifted upward relative to the modal low-degree group. This pattern links the output shift to the graph feature used for offset assignment as well as to the aggregate change in candidate count.

Supplementary controls narrow the interpretation further. On the canonical \(N=70\)--150 mappings, degree and physical-chain length were positively correlated in all 120 graph instances (mean \(\rho=0.67\)). For the 118 instances with certified occurrences under both protocols, the degree--shift partial rank correlation after accounting for chain length remained positive in 117 instances (mean \(\rho=0.24\)), whereas the complementary chain-length association after accounting for degree averaged \(-0.11\) (Supplementary Fig.~S5). In the matched shuffle control, the original degree-aligned assignment also produced a positive degree--shift correlation relative to shuffled assignments with the same offset multiset in all 20 graphs (mean \(\rho=0.17\)) (Supplementary Fig.~S6). A separate random 3-regular ensemble reversed the sparse-BA direction, favoring QA-Base, when the rule reduced to a uniform offset (Supplementary Table~S5 and Fig.~S9). A 10-instance connected ER control targeting nominal mean degree four also favored QA-Base in certified-support size, indicating that approximate mean-degree matching alone was insufficient to transfer the sparse-BA result (Supplementary Fig.~S10). Taken together, these controls provide evidence beyond the offset multiset alone, but the 3-regular comparison changes the graph ensemble as well as the offset pattern and therefore is a transfer boundary rather than a same-graph, mean-matched uniform-offset control.

Together, the matched-shuffle and partial-rank controls associate the support shift with offset placement after accounting for additive monotone chain-length effects. Time-resolved device studies could test nonlinear chain-length interactions and the corresponding dynamics.

An embedding-free closed-system model likewise produced placement-dependent redistribution and an intermediate-budget discovery increase, establishing sufficiency within the specified Hamiltonian. This parallels the hardware shuffle control but does not establish correspondence to the QPU's open-system dynamics (Supplementary Figs.~S12 and S13).

\subsection{Finite-budget support and diversity}

Under an independent-draw approximation, let \(P(x)\) be the unconditional per-read probability of certified optimum \(x\) in manifold \(\mathcal{M}\), so that \(\sum_{x\in\mathcal{M}}P(x)\) equals the certified yield. The expected number of distinct certified optima observed in \(R\) raw reads is
\[
\mathbb{E}[U_{R}]=\sum_{x\in\mathcal{M}}\left[1-(1-P(x))^{R}\right].
\]
This expression separates two contributors to certified-support size: the probability of producing a certified occurrence and the diversity conditional on producing a certified optimum. In the 50,000-read diagnostic, DC-OSQA had both higher mean certified yield and higher exact expected support after conditioning on the same number of certified draws (exact rarefaction; Supplementary Fig.~S11(a,b)). Both components therefore contributed to the observed support expansion and remain properties of the observed sample.

The Hamming-family analysis shows that the observed candidates occupy multiple mutually separated groups under the chosen threshold. Complete linkage enforces that every pair within a family satisfies the threshold. The positive paired difference persists over the tested sensitivity grid \(\tau/N=0.00\)--0.20 (Supplementary Fig.~S11(c)), but family count still depends mechanically on observed support size. We therefore interpret family count as a secondary, threshold-dependent structural summary of observed support.

\subsection{Scope and limitations}

The strongest evidence concerns sparse BA \(m=2\) MVC instances on one QPU solver. Denser BA controls indicate that the favorable offset scale varies with graph density (Supplementary Fig.~S7). The random 3-regular and connected-ER reversals delimit transfer to other graph ensembles (Supplementary Figs.~S9 and S10). The ER graphs were selected for connectedness and matched the BA ensemble by nominal mean degree, while realized edge count, degree-distribution shape, and certified yield also differed. Identifying the structural source of the reversal therefore requires separately matched controls. Within the primary ensemble, \(\alpha=0.1\) serves as an empirical operating point.

The simulated-annealing comparison provides support-recovery context. The paired QPU design and supplementary embedding and chain diagnostics address several hardware-dependent factors within the tested acquisition; broader hardware and repeated-batch studies would test transfer beyond this setting.

\subsection{Outlook}

The results motivate treating anneal offsets as problem-derived sampling controls rather than only as generic calibration parameters. In MVC, max-normalized degree provides a transparent local signal because it counts direct constraint participation. Other constrained optimization problems may admit analogous rules based on variable participation or influence, but those rules require independent validation.

A practical next step is to test whether broader certified support improves a downstream task. Certified candidates could seed classical local search or other hybrid procedures, following the broader principle of using quantum-annealing samples to identify useful solution structure \cite{pang2021potential}. Stronger transfer tests should also include additional graph families, real network instances where feasible, matched homogeneous-offset controls on the same BA graphs, and repeated hardware batches with complete calibration and execution metadata.

\section{Conclusion}
We introduced DC-OSQA, an MVC-specific static anneal-offset rule derived from max-normalized vertex degree, and evaluated it as a control of finite-read sampling for Minimum Vertex Cover. On sparse BA \(m=2\) graphs with \(N=70\)--150, DC-OSQA increased both the observed number of distinct certified candidates and the number of complete-linkage Hamming families relative to the matched QA-Base condition. Concentration, degree-resolved, regenerated-embedding, and shuffled-assignment diagnostics associate this change with lower empirical concentration and degree-aligned redistribution, while the ER and 3-regular controls delimit transfer beyond the primary ensemble.

Within the tested sparse BA \(m=2\) ensemble and hardware setting, DC-OSQA shows that an application-specific assignment of local annealing controls can change which certified optima are observed within a fixed QPU read budget.

\section*{Acknowledgements}
This work was supported by the Korean ARPA-H Project through the Korea Health Industry Development Institute (KHIDI), funded by the Ministry of Health and Welfare (RS-2025-25456722); the Brain Pool Plus Fellowship Program, funded by the Ministry of Science and ICT (RS-2025-25427881); and the ``Creation of the quantum information science R\&D ecosystem (based on human resources)'' through the National Research Foundation of Korea (NRF), funded by the Korean government (RS-2023-NR057243). J.K. was funded by the Department of Quantum Information, Yonsei University.

\section*{Competing interests}
The authors declare no competing interests.

\section*{Data availability statement}
The data and analysis code supporting this study are publicly available in the GitHub repository \cite{kang2026dcosqa_data}.

{\footnotesize
\bibliographystyle{unsrt}
\bibliography{references}
}

\clearpage


\section*{Supplementary material}

# Supplementary information

**Article:** Degree-conditioned anneal offsets reshape certified-optimum sampling for minimum vertex cover

**Authors:** Jiwan Kang and Namshik Han

**Description:** Additional definitions, control experiments, sensitivity analyses, and finite-budget diagnostics supporting the main article.

## Supplementary information overview

The controls address distinct alternatives to the primary comparison. Shuffled assignments preserve the offset multiset while changing its placement; regenerated embeddings test dependence on one cached embedding; offset-scale and graph-density studies test operating-point sensitivity; and random 3-regular graphs provide a degree-regular boundary case. Vertex-level mappings from the canonical comparison are used to examine degree–chain-length confounding, and connected ER graphs targeting nominal mean degree four provide an additional topology-transfer control. Simulated-annealing (SA) recovery, QPU read-budget growth, and support-overlap analyses describe finite-budget relationships under the tested implementations. Finally, an idealized closed-system state-vector calculation tests whether degree-aligned local schedules are sufficient to produce placement-dependent redistribution within the specified small-system Hamiltonian. All QPU experiments used D-Wave's `Advantage2_system1` solver.

## Contents



# S1 Data curation and QPU implementation

## S1.1 Certified-candidate validation and coverage

Protocol labels were standardized before analysis, and redundant exports of the same source records were removed using source provenance. Identical bitstrings from distinct runs or with separate occurrence counts were retained as sampling evidence; repeated representations within a protocol–instance group were consolidated before support and concentration calculations.

A sample was retained as a certified candidate only when its cardinality equaled the independently computed minimum cover size $K_{\text{exact}}$ and every vertex-cover constraint was satisfied. Cardinality-matched but infeasible bitstrings were excluded. Protocol–instance pairs with no certified candidates were retained with zero unique candidates and zero Hamming families. Each canonical QPU protocol–instance pair used 1000 reads. The canonical graph-size conditions used 10 graph instances for each size in $N = 20$–$60$ and 20 for each size in $N = 70$–$150$; the $N = 200$ diagnostic was analyzed separately. Two canonical zero-yield groups occurred, both for QA-Base at BA $m = 2$: $N = 120$, instance 4 and $N = 150$, instance 15. In each case all 1000 reads were retained in the denominator, but none met both certification conditions.

## S1.2 Embeddings, schedules, and chain handling

Logical QUBOs were embedded using `minorminer.find_embedding` [1, 2]. For each canonical graph instance, QA-Base and DC-OSQA shared one physical embedding held fixed through `FixedEmbeddingComposite`; the retained vertex-to-chain mappings are included in the released data and support the diagnostic in Sec. S5.4. The canonical QA-Base schedule was a monotonic $20\,\mu$s anneal without node-dependent offsets or an explicit pause. DC-OSQA used the same base anneal with $\delta_i = -\alpha d_i / d_{\max}$, where $\alpha = 0.1$ in the primary experiment.

The logical offset for variable $i$ was assigned to every physical qubit in its embedded chain. Requested values were checked against the QPU-reported `anneal_offset_ranges` [3]. No explicit chain strength was supplied; the embedding composite's default handling was used consistently within matched comparisons. Broken chains were decoded by majority vote, and per-sample chain-break fractions were retained as diagnostics. Vertex-level physical-chain mappings were retained for both the canonical comparison and the 100 regenerated embeddings.

Betweenness-centrality OSQA (BC-OSQA) used maximum-normalized NetworkX betweenness centrality [4, 5] with the same negative-offset convention. It is a graph-topological comparator, whereas degree directly counts the MVC edge constraints incident on a variable.

# S2 Hamming families and observed concentration

## S2.1 Complete-linkage Hamming families

Complete-linkage Hamming families were constructed from the unique certified candidates using the Hamming-distance and clustering definitions in the main-text Methods [6]. The primary threshold was $\tau = 0.1N$; cutting the complete-linkage dendrogram at this threshold ensures that every pair of candidates within a final family differs by at most $\tau$ bits. Family count is therefore interpreted as a threshold-dependent summary of the observed support.

## S2.2 Observed certified-sample concentration

For $M$ distinct certified candidates with $C$ total certified occurrences, let $p_j$ denote candidate $j$'s empirical occurrence probability. Table S1 reports the four concentration diagnostics defined in the main-text Methods together with two additional summaries: duplicate rate $1 - M/C$ and the top-1 share, the occurrence mass of the single most frequent observed candidate. The top-decile share used in main-text Figure 3 sums the mass of the $\lceil 0.1M \rceil$ most frequent candidates and differs from the fixed top-10-candidate share used only in the 50,000-read table below. All concentration metrics are conditional on the observed certified occurrences.

Table S1: Observed certified-sample concentration for BA $m = 2$, $N = 70$–150. The Direction column reports the DC-OSQA mean relative to the QA-Base mean; $p$-values are from two-sided paired Wilcoxon tests.

| Metric | Pairs | DC mean | QA mean | $p$-value | Direction |
|---|---|---|---|---|---|
| Duplicate rate | 118 | 0.38 | 0.60 | $3.16 \times 10^{-16}$ | lower |
| Normalized entropy | 118 | 0.95 | 0.84 | $3.32 \times 10^{-19}$ | higher |
| Simpson support fraction | 118 | 0.69 | 0.41 | $3.69 \times 10^{-17}$ | higher |
| Top-1 share | 118 | 0.04 | 0.15 | $7.84 \times 10^{-20}$ | lower |
| Top-decile share | 118 | 0.27 | 0.47 | $3.16 \times 10^{-19}$ | lower |
| Gini coefficient | 118 | 0.27 | 0.48 | $3.77 \times 10^{-17}$ | lower |

# S3 Larger-size diagnostic at $N = 200$

The $N = 200$ data were treated as an extension rather than included in the primary $N = 70$–150 claim. DC-OSQA retained positive paired differences for both outcomes; BC-OSQA showed a positive family-count difference but a less consistent certified-support-size difference (Table S2). For certified-support size, the $N = 200$ paired test gave $p = 3.62 \times 10^{-2}$. Because this extension contains a single additional size, it was not used to infer scaling with $N$.

# S4 Simulated-annealing recovery and support overlap

Simulated annealing [7] served as a recovery and set-overlap probe on 20 BA $m = 2$, $N = 150$ instances. The analyses used a separately retained 1,000-sample SA-Max acquisition and

Table S2: Paired $N = 200$ diagnostic over 20 matched graph instances. W/L/T denotes wins, losses, and ties for the listed protocol relative to QA-Base. Mean gain is protocol minus QA-Base, and $p$-values are from two-sided paired Wilcoxon tests.

| Metric | Protocol | W/L/T | Protocol mean | QA mean | Mean gain | $p$ |
|---|---|---|---|---|---|---|
| Unique certified | DC-OSQA | 16/4/0 | 387.35 | 263.25 | 124.10 | $3.62 \times 10^{-2}$ |
| Unique certified | BC-OSQA | 14/6/0 | 332.60 | 263.25 | 69.35 | $2.03 \times 10^{-1}$ |
| Hamming families | DC-OSQA | 18/2/0 | 61.70 | 17.10 | 44.60 | $8.31 \times 10^{-4}$ |
| Hamming families | BC-OSQA | 16/3/1 | 42.45 | 17.10 | 25.35 | $5.13 \times 10^{-3}$ |

independent recovery acquisitions of 10,000 and 50,000 samples; all used 10,000 sweeps and an inverse-temperature range $\beta \in [0.1, 5.0]$. Across the two recovery acquisitions, the mean unique certified-candidate count increased from 4439.45 at 10,000 samples to 19,040.15 at 50,000 samples, with an increase on every instance (Wilcoxon $p = 1.91 \times 10^{-6}$). The released candidate tables and analysis scripts reproduce the reported recovery and set-overlap summaries.

For observed DC-OSQA support $D$ and SA support $S$, we use

$$\text{Jaccard}(D, S) = \frac{|D \cap S|}{|D \cup S|}, \qquad \text{DC-support recovery by SA} = \frac{|D \cap S|}{|D|}, \tag{S1}$$

$$\text{union gain} = \frac{|D \cup S|}{\max(|D|, |S|)}. \tag{S2}$$

A union gain above one indicates that the union contains more certified candidates than the larger individual observed support.

We evaluated two distinct finite-budget questions. First, SA recovered mean fractions 0.12 and 0.31 of the canonical 1,000-read DC-OSQA support at 10,000 and 50,000 samples, respectively (Fig. S1(a)). Second, DC-OSQA and SA were compared at matched output budgets of 1,000, 10,000, and 50,000 samples per sampler (Table S3; Fig. S1(b)). The 1,000- and 10,000-read DC-OSQA summaries average 100 without-replacement subsamples from each 50,000-read stream; the 50,000-read comparison uses the complete stream.

Table S3: Matched-output-budget DC-OSQA/SA support decomposition at $N = 150$, averaged over 20 graph instances. Support columns give mean certified-support sizes; the remaining columns give the mean intersection size and mean fractions of the union. DC-OSQA values at 1,000 and 10,000 reads average 100 without-replacement subsamples per instance from the 50,000-read stream.

| Output budget per sampler | DC support mean | SA support mean | Shared mean | DC-only (% union) | Shared (% union) | SA-only (% union) |
|---|---|---|---|---|---|---|
| 1,000 | 565.31 | 477.45 | 7.89 | 48.44 | 0.79 | 50.77 |
| 10,000 | 4303.82 | 4439.45 | 313.26 | 39.58 | 5.31 | 55.11 |
| 50,000 | 15929.55 | 19040.15 | 2226.25 | 28.51 | 14.20 | 57.29 |

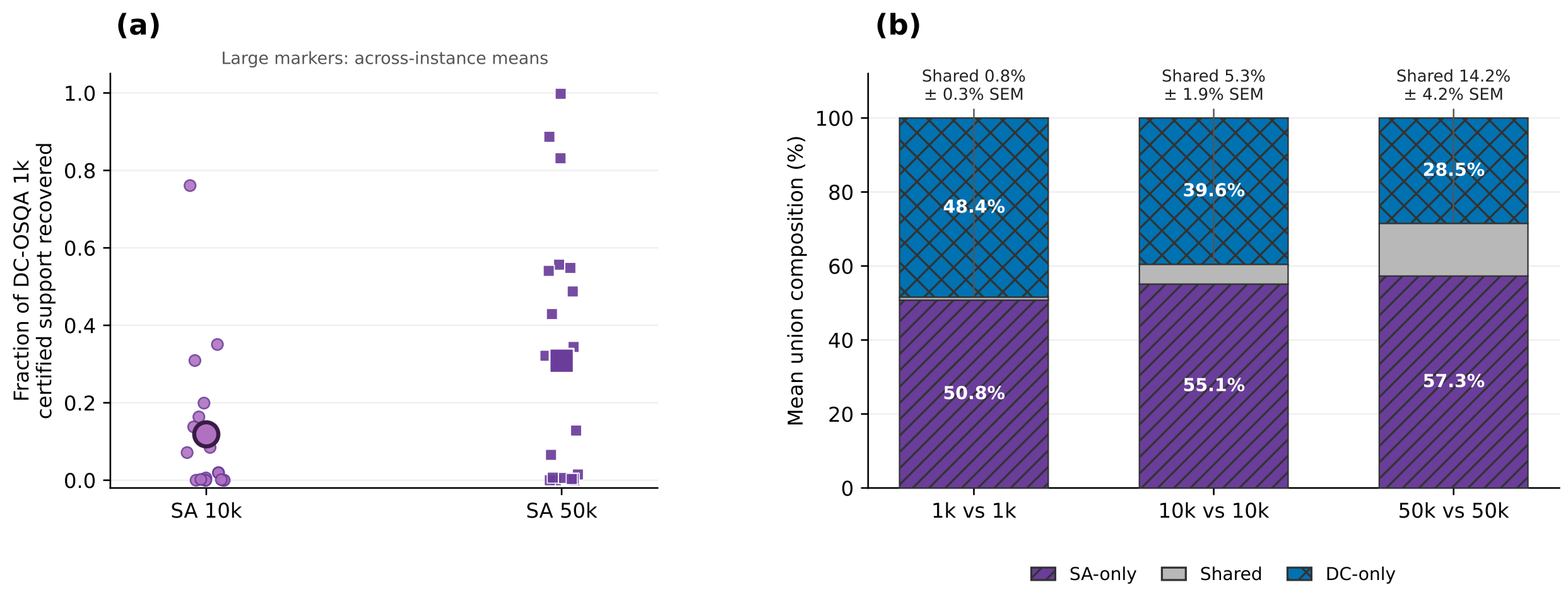


Figure S1: Finite-budget complementarity of DC-OSQA and SA at $N = 150$. (a) Fraction of the canonical 1,000-read DC-OSQA support recovered by 10,000- and 50,000-sample SA runs; small markers show individual instances and large markers show across-instance means. (b) Matched-output-budget union decomposition into SA-only, shared, and DC-only fractions; bars show mean fractions across 20 instances, and annotations give the mean shared fraction ± standard error of the mean (s.e.m.). The 1,000-sample SA comparison uses the separately retained SA-Max dataset, whereas the 10,000- and 50,000-sample comparisons use the corresponding SA recovery datasets. DC-OSQA values at 1,000 and 10,000 reads average 100 without-replacement subsamples from each 50,000-read stream.

# S5 Control and sensitivity analyses

## S5.1 Betweenness-centrality comparator

BC-OSQA produced positive mean gains over QA-Base, but DC-OSQA showed larger mean paired gains at every tested size for both outcomes. These results motivate retaining degree as the primary MVC-specific signal in the present comparison (Fig. S2).

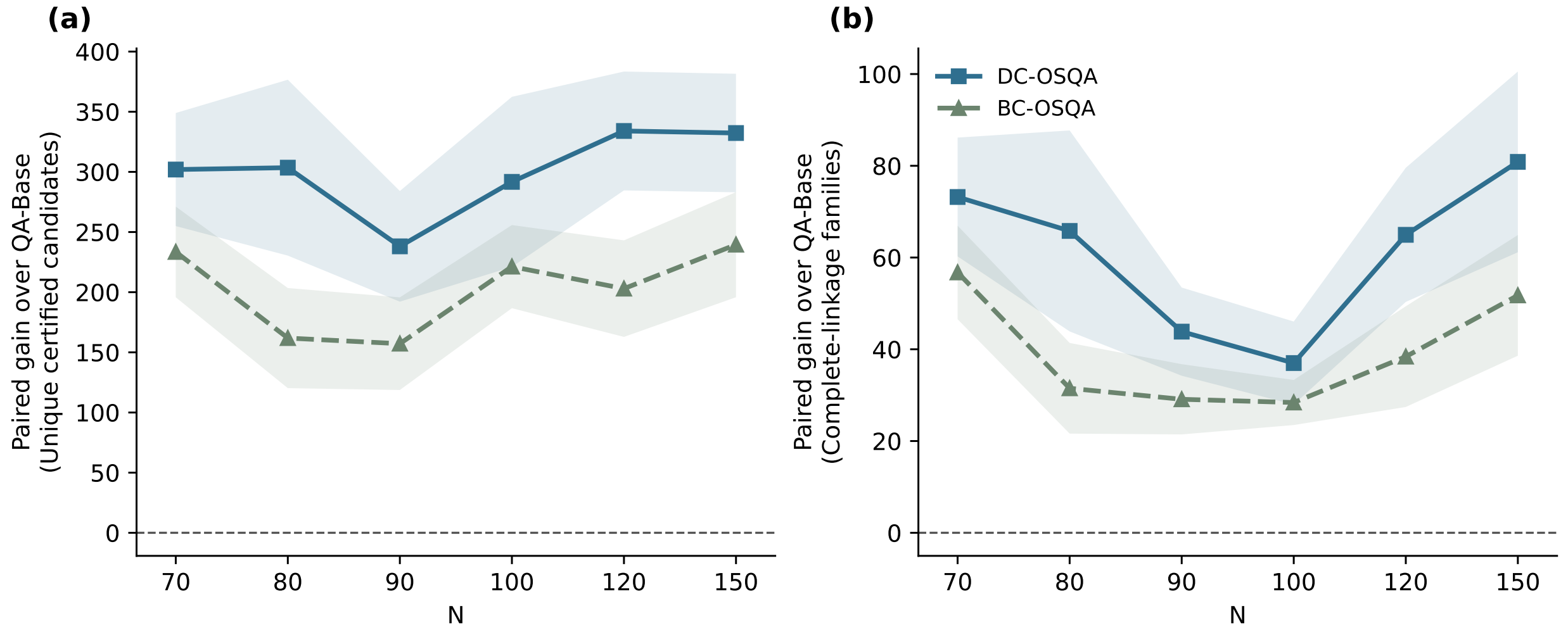


Figure S2: Degree- and betweenness-centrality offset schedules on BA $m = 2$, $N = 70$–$150$. Curves show mean paired gains over QA-Base in (a) unique certified candidates and (b) complete-linkage Hamming families; bands show the standard error across graph instances.

## S5.2 Offset-scale sensitivity

DC-OSQA was evaluated at $\alpha \in \{0.05, 0.1, 0.2\}$ on matched BA $m = 2$ instances at $N \in \{70, 90, 120, 150\}$. The canonical $\alpha = 0.1$ setting was fixed across graph sizes and was not tuned by instance. Both outcomes varied with $\alpha$, demonstrating sensitivity to the offset scale (Fig. S3).

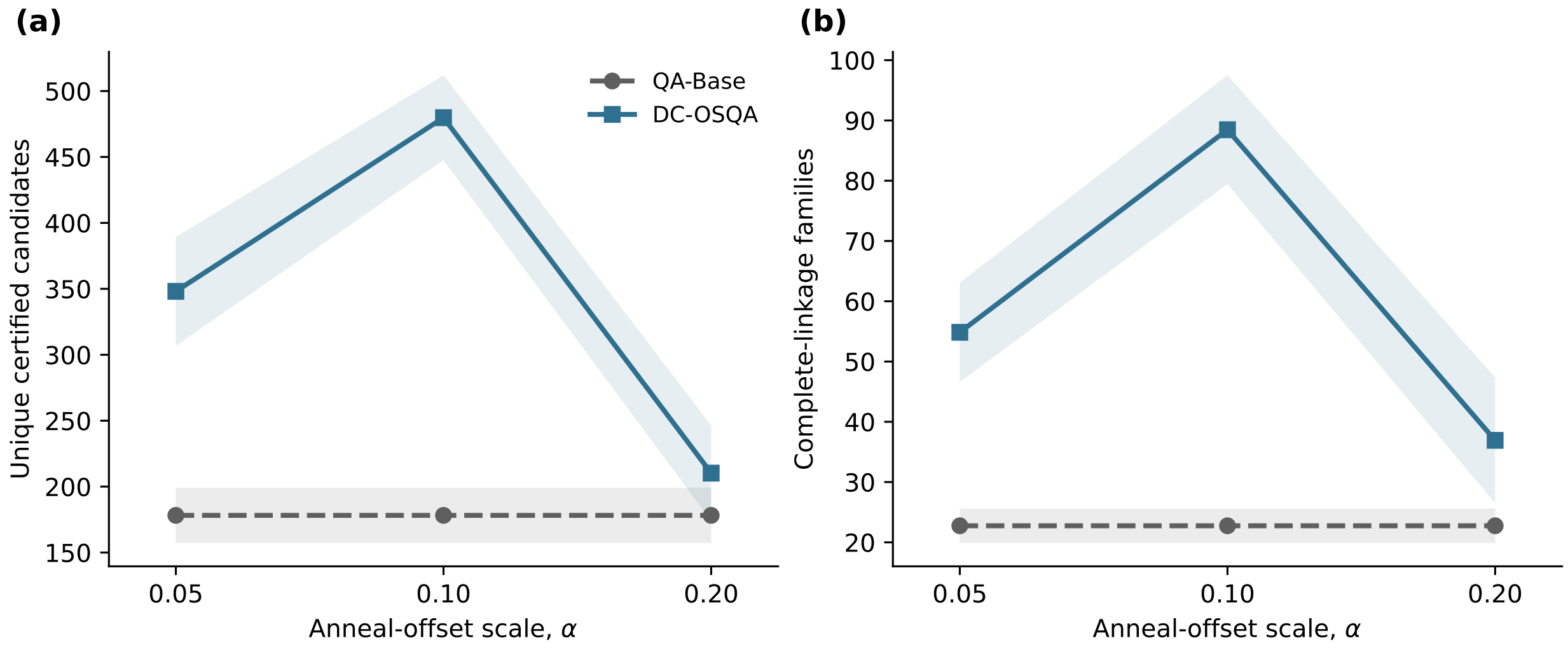


Figure S3: DC-OSQA offset-scale sensitivity on matched BA $m = 2$ instances. (a) Mean unique certified candidates and (b) mean complete-linkage Hamming families as a function of the offset scale $\alpha$; dashed gray lines show the matched QA-Base baseline, and bands denote the standard error across graph instances.

## S5.3 Embedding-seed sensitivity

Five new minor embeddings were generated for each of the 20 canonical BA $m = 2$, $N = 150$ graphs. QA-Base and DC-OSQA were paired on the same regenerated embedding, giving 100 graph–embedding pairs. DC-OSQA won 84 pairs in unique certified candidates and 92 pairs in Hamming families. After averaging over the five embeddings per graph, all 20 graph-level gains were positive for both outcomes. Graph-level means were 487.11 versus 253.06 unique candidates and 85.81 versus 27.61 families (both $p = 1.91 \times 10^{-6}$). Individual embeddings sometimes favored QA-Base (Fig. S4).

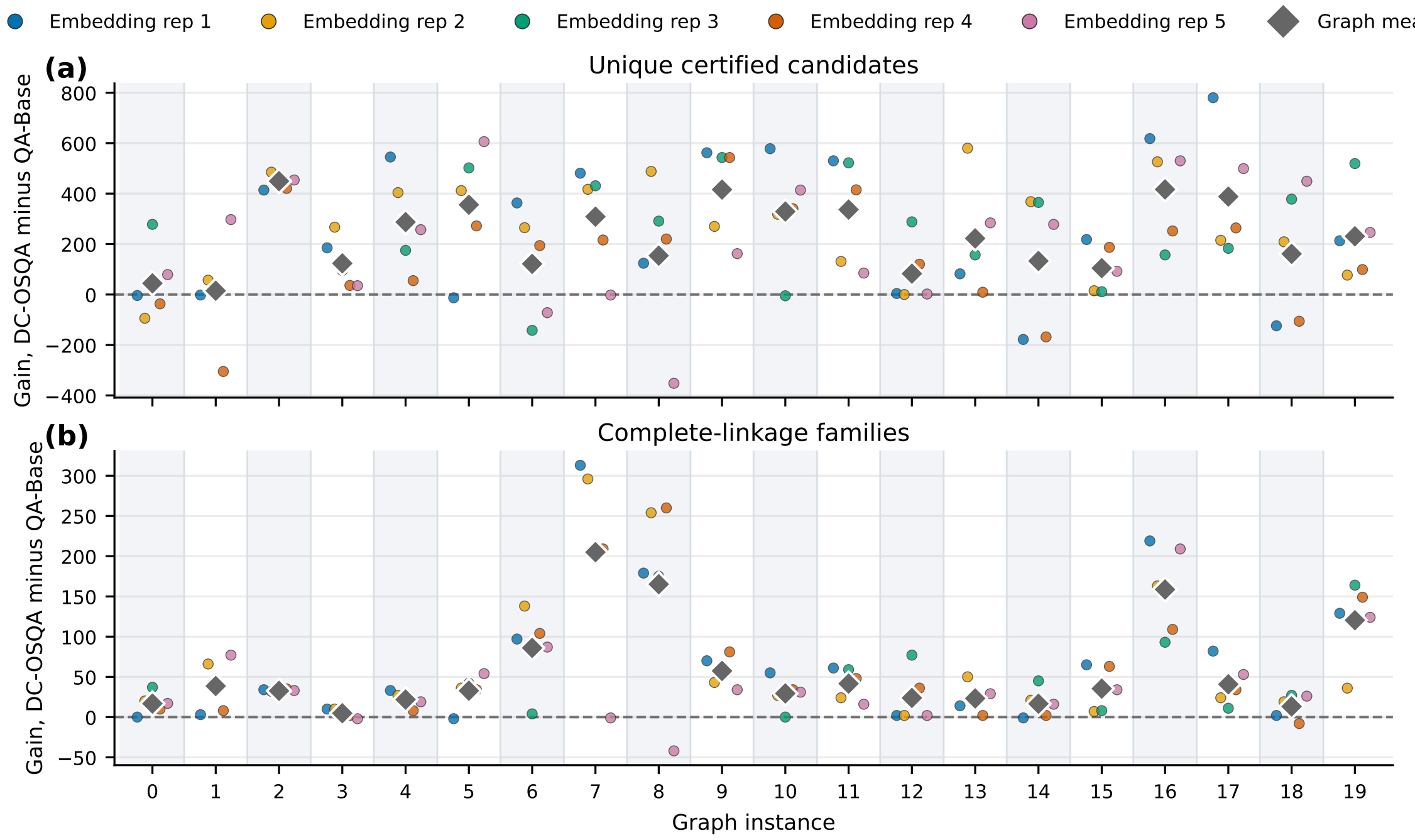


Figure S4: Embedding-seed sensitivity at $N = 150$. Paired DC-OSQA minus QA-Base gains in (a) unique certified candidates and (b) complete-linkage Hamming families. Small points show individual regenerated embeddings, diamonds show graph-level means over the five embeddings, and the dashed line marks zero gain.

## S5.4 Vertex-level chain-length diagnostic

The canonical release contains the physical chain assigned to every logical vertex for the 120 BA $m = 2$, $N = 70$–$150$ graph instances in the main larger-size analysis. Chain length ranged from 1 to 9 physical qubits (overall mean 1.52). For each graph, we first computed the within-instance Spearman correlation between logical-vertex degree and chain length. All 120 correlations were positive, with mean $\rho = 0.67$, median 0.68, and range 0.42–0.83. The strong association between degree and physical chain length motivated the adjusted analysis below.

For each protocol, vertex $i$'s inclusion probability $p_i$ was computed over certified occurrences, retaining occurrence multiplicities. We then defined the paired inclusion shift $\Delta p_i = p_i^{\mathrm{DC}} - p_i^{\mathrm{QA}}$. Both probabilities were defined for 118 of the 120 graph instances; the two QA-Base zero-yield instances were excluded from this conditional diagnostic.

To separate the two monotone associations descriptively, degree, chain length, and $\Delta p_i$ were converted to within-pair ranks. The partial degree association was the Pearson correlation between residuals from separately regressing ranked degree and ranked $\Delta p_i$ on ranked chain length. We denote this statistic by $\rho_{\Delta p, d\cdot\ell}$; the complementary statistic $\rho_{\Delta p, \ell\cdot d}$ exchanges degree $d$ and chain length $\ell$.

The unadjusted degree–shift correlation was positive in all 118 instances (mean $\rho = 0.23$). After rank-residualizing chain length, the degree–shift association remained positive in 117 of 118 instances, with mean partial $\rho = 0.24$, median 0.25, and range $-0.16$ to 0.52. The unadjusted chain-length–shift correlation was positive in 95 instances, negative in 19, and zero in four (mean $\rho = 0.08$), whereas the chain-length association after rank-residualizing

degree was negative in 104 instances and averaged $-0.11$. The residualization addresses additive monotone associations; nonlinear chain-length-by-offset interactions remain outside this diagnostic (Fig. S5).

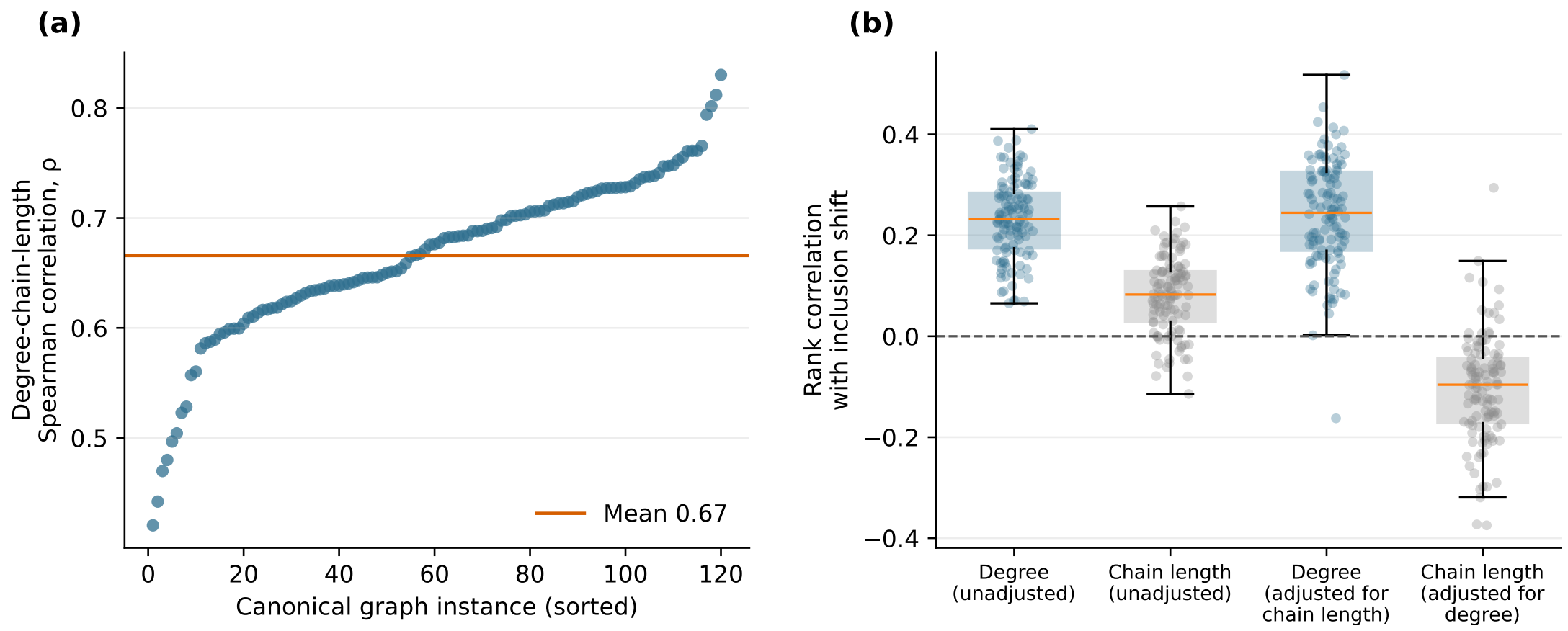


Figure S5: Vertex-level chain-length diagnostic for the canonical BA $m = 2$, $N = 70$–150 comparison. (a) Within-instance Spearman correlations between logical degree and physical chain length for 120 graph mappings; the orange line marks the mean. (b) Unadjusted and rank-residualized associations with the occurrence-weighted inclusion shift (DC-OSQA minus QA-Base) across the 118 instances for which both conditional probabilities were defined. Boxes show medians and interquartile ranges, whiskers extend to 1.5 times the interquartile range, and points show graph instances. "Degree (adjusted for chain length)" denotes $\rho_{\Delta p, d \cdot \ell}$, and "chain length (adjusted for degree)" denotes $\rho_{\Delta p, \ell \cdot d}$.

## S5.5 Matched offset-assignment control

For each of 20 BA $m = 2$, $N = 150$ graphs, the degree-aligned assignment was compared with 20 hardware-valid random reassignments of the same logical offset multiset on a fixed fresh embedding. This preserves the graph, QUBO, embedding, read budget, certification rule, and offset values while changing their logical-variable placement. Paired support summaries are reported in Table S4.

Table S4: Matched 20-shuffle control. W/L gives DC-OSQA wins and shuffle-summary wins over 20 graphs. Mean, median, and best of 20 summarize the shuffled assignments within each graph; mean gain is DC-OSQA minus the corresponding shuffle summary, and $p$-values are from two-sided paired Wilcoxon tests.

| Metric | Shuffle summary | DC mean | Shuffle mean | Mean gain | W/L | $p$ |
|---|---|---|---|---|---|---|
| Unique certified | Mean | 561.40 | 362.65 | 198.76 | 19/1 | $4.8 \times 10^{-5}$ |
| Unique certified | Median | 561.40 | 365.83 | 195.58 | 19/1 | $6.3 \times 10^{-5}$ |
| Unique certified | Best of 20 | 561.40 | 513.55 | 47.85 | 13/7 | 0.12 |
| Hamming families | Mean | 96.35 | 45.91 | 50.44 | 20/0 | $2.0 \times 10^{-6}$ |
| Hamming families | Median | 96.35 | 45.95 | 50.40 | 20/0 | $8.8 \times 10^{-5}$ |
| Hamming families | Best of 20 | 96.35 | 68.90 | 27.45 | 17/3 | $6.04 \times 10^{-3}$ |

For vertex $i$, let $p_i^{\text{orig}}$ be its occurrence-weighted inclusion probability among certified candidates under the degree-aligned assignment and $\bar{p}_i^{\text{shuf}}$ the mean over 20 shuffles. The placement-sensitive shift is

$$\Delta p_i^{\text{orig}-\text{shuf}} = p_i^{\text{orig}} - \bar{p}_i^{\text{shuf}}. \quad \text{(S3)}$$

For panel (b), vertices were grouped by actual degree as low ($d \leq 2$), middle ($3 \leq d \leq 4$), and high ($d \geq 5$). This is the same fixed BA $m = 2$ definition used in the canonical $N = 70$–150 analysis; it was not recalculated for the $N = 150$ control. The ranges isolate the minimum and modal degree, degrees up to the ensemble mean, and the above-mean tail while retaining all tied-degree vertices in the same group. Across the 20 graphs, the groups contained on average 75.0, 43.4, and 31.6 vertices, respectively. Their mean placement-sensitive shifts were −0.03, +0.05, and +0.01. The within-graph Spearman correlation between $\Delta p_i^{\text{orig-shuf}}$ and degree was positive in all 20 graphs (mean $\rho = 0.17$, range 0.04–0.31) (Fig. S6).

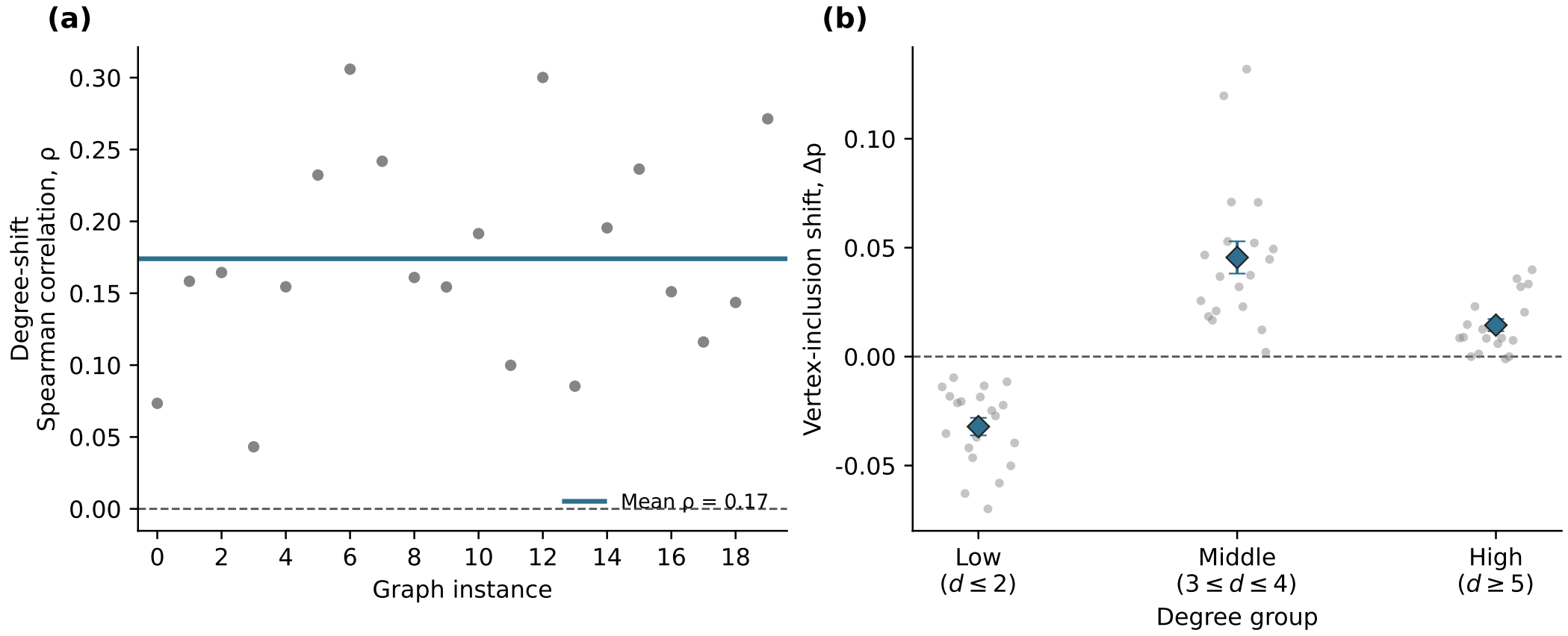


Figure S6: Degree alignment in the matched shuffle control. (a) Within-graph Spearman correlations between degree and $\Delta p_i^{\text{orig-shuf}}$; the blue horizontal line marks the across-graph mean. (b) Vertices grouped by actual degree as $d \leq 2$, $3 \leq d \leq 4$, and $d \geq 5$; gray points are graph-level group means, and blue diamonds show across-graph means with s.e.m. error bars. Group means are not connected because the three ranges are categorical and unequal in degree width.

## S5.6 Denser BA graphs

BA $m = 3$ and $m = 4$ controls used $N \in \{70, 90, 120, 150\}$, 10 graphs per size, and $\alpha \in \{0.01, 0.025, 0.05, 0.10\}$. BA $m = 3$ retained positive gains over much of the range but often favored $\alpha = 0.05$ for $N \geq 90$. For BA $m = 4$, gains weakened at larger sizes and $\alpha = 0.1$ became detrimental at $N = 120$ and 150. Thus, the favorable offset scale varied with graph density (Fig. S7).

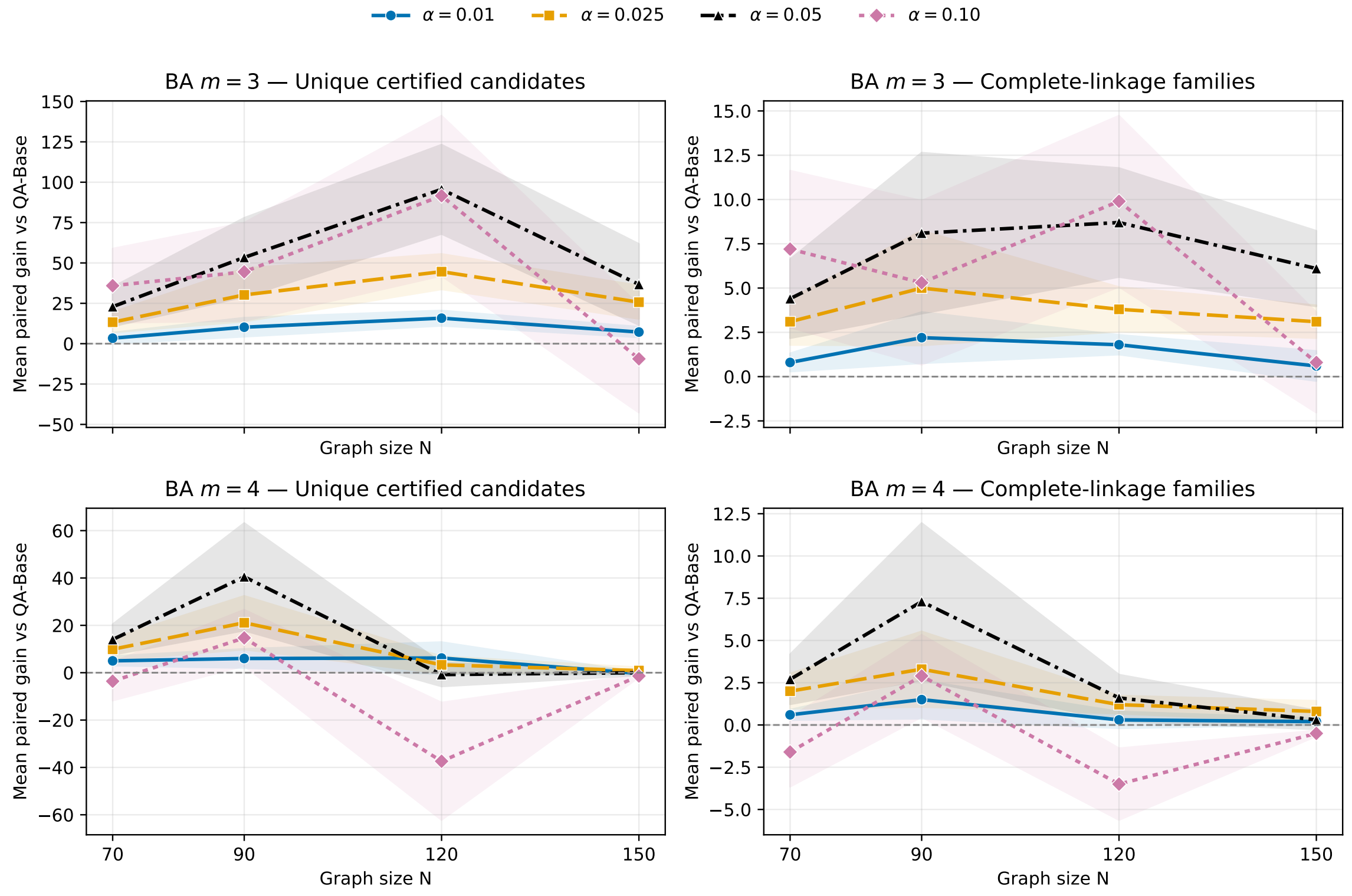


Figure S7: Offset-scale controls for BA $m = 3$ (top row) and $m = 4$ (bottom row). Curves show mean paired DC-OSQA minus QA-Base gains in unique certified candidates (left column) and complete-linkage Hamming families (right column) at each offset scale $\alpha$; bands show the standard error across 10 graphs, and the dashed horizontal line marks zero gain.

## S5.7 Degree-resolved composition shift

Across 118 canonical matched comparisons for which both conditional summaries were defined, vertices were grouped by the fixed BA $m = 2$ ranges low ($d \leq 2$), middle ($3 \leq d \leq 4$), and high ($d \geq 5$). The thresholds were held constant across $N = 70$–150; over these sizes the groups represented 47.4–50.0%, 28.5–31.1%, and 21.1–22.4% of vertices, respectively. Shifts (DC-OSQA minus QA-Base) under equal weighting of distinct observed candidates were $-0.04$, $+0.05$, and $+0.02$. Occurrence-weighted shifts were $-0.05$, $+0.06$, and $+0.03$. The actual-degree grouping keeps tied-degree vertices together and shows that the mean increase is largest for $d = 3$–4 rather than in the highest-degree tail (Fig. S8).

As a convergent check, the regenerated-embedding data gave 93 graph–embedding pairs in which both protocols produced certified occurrences. Within-pair Spearman correlations between degree and $p_i^{\mathrm{DC}} - p_i^{\mathrm{QA}}$ were positive in all 93 pairs (mean $\rho = 0.21$, range 0.03–0.40). This statistic is undefined for the seven pairs in which one protocol had zero certified occurrences.

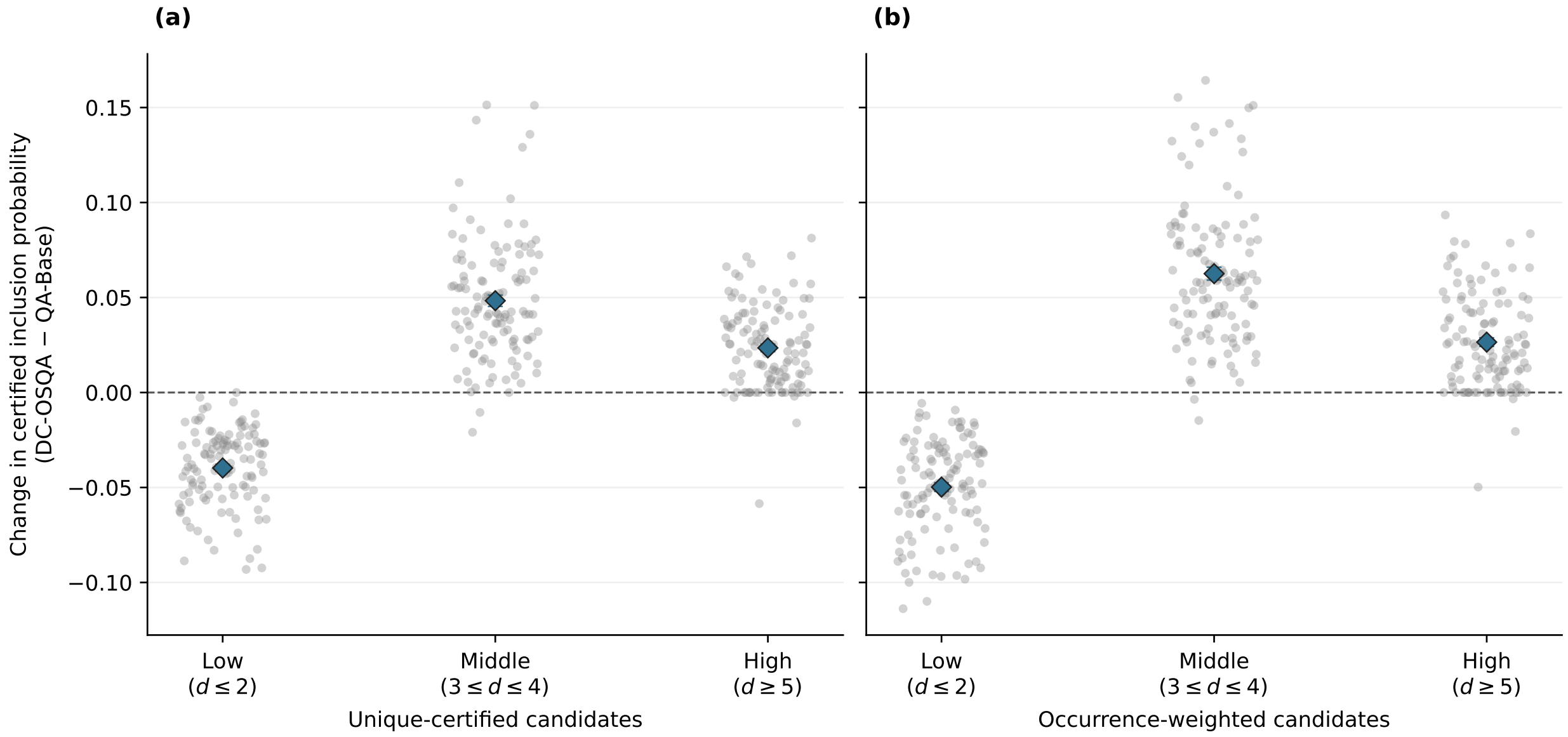


Figure S8: Degree-associated inclusion shifts for BA $m = 2$, $N = 70$–150. Panel (a) weights distinct observed candidates equally; panel (b) weights their observed occurrences. Vertices are grouped by actual degree as $d \leq 2$, $3 \leq d \leq 4$, and $d \geq 5$. Gray points show graph-instance group means, and blue diamonds show across-instance means with s.e.m. error bars. Group means are not connected because the three ranges are categorical and unequal in degree width.

## S5.8 Random 3-regular boundary control

Twenty random 3-regular graphs with $N = 150$ were tested at 1,000 reads per protocol. Because every vertex has $d_i = 3$, DC-OSQA reduces to the uniform shift $\delta_i = -0.1$. The result reversed the sparse-BA direction: QA-Base exceeded DC-OSQA in 19 of 20 instances for unique certified candidates and certified occurrences, and in 18 of 20 instances for Hamming families, with one tie (Table S5; Fig. S9). This changes both graph ensemble and offset pattern, so it is a transfer boundary rather than a same-graph, mean-matched uniform-offset control.

Table S5: Random 3-regular boundary control. Mean and median gains are DC-OSQA minus QA-Base; negative gains favor QA-Base.

| Metric | QA mean | DC mean | Mean gain | Median gain | DC/QA wins |
|---|---|---|---|---|---|
| Unique certified | 299.35 | 180.75 | −118.60 | −68.5 | 1/19 |
| Certified occurrences | 438.80 | 258.95 | −179.85 | −152.5 | 1/19 |
| Hamming families | 27.70 | 19.35 | −8.35 | −3.0 | 1/18 (1 tie) |

The mean feasible-occurrence rates were 99.82% for QA-Base and 99.94% for the uniform-offset protocol, while certified yields were 0.44 and 0.26. The feasible-occurrence rate is the fraction of reads that decode to a feasible vertex cover of any cardinality; the certified yield is $Y = C/R$, where $R$ is the read budget. Thus, most noncertified reads in this control were feasible covers with cardinality above $K_{\text{exact}}$: feasibility and minimum-cover recovery measure different properties of the output. No sample in either protocol had a chain-break fraction above 1%. The reversal therefore coincided with a lower probability of producing a

certified optimum, not with a measured feasibility or large-chain-break imbalance.

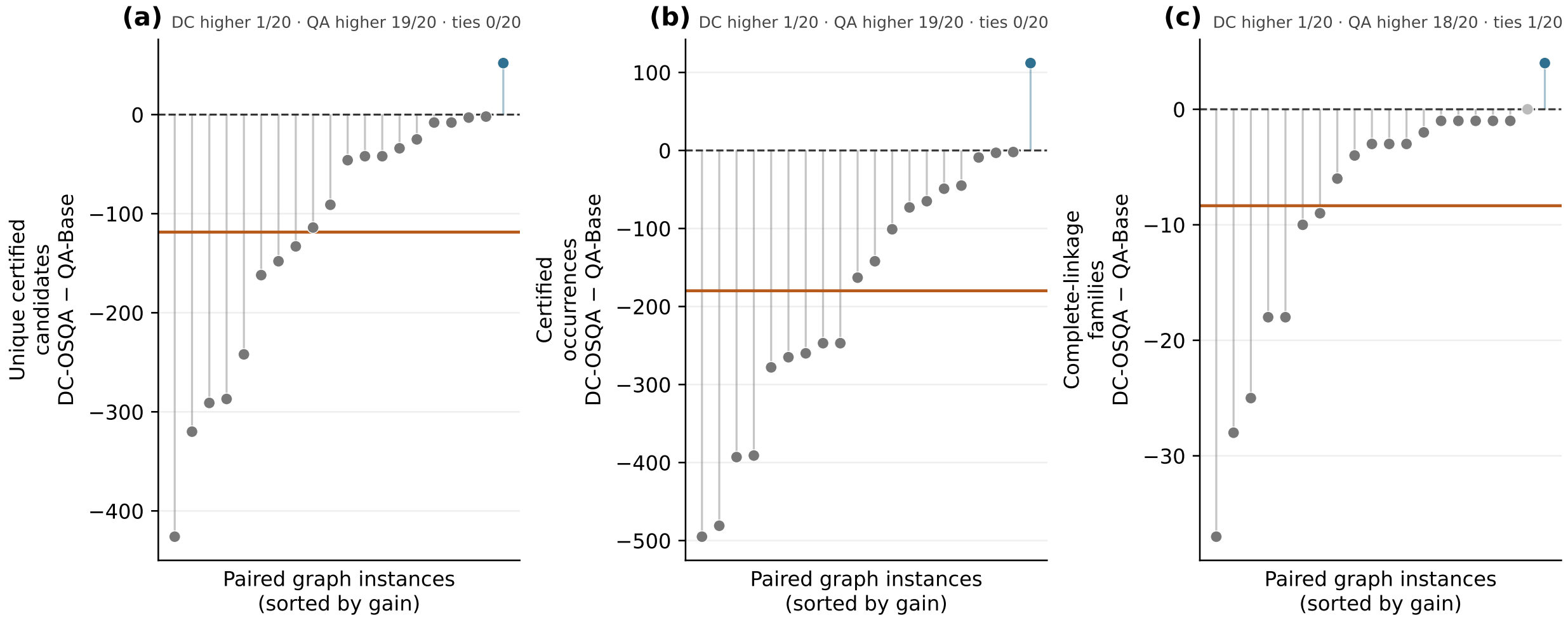


Figure S9: Random 3-regular boundary control at $N = 150$. Panels show instance-paired DC-OSQA minus QA-Base gains in (a) unique certified candidates, (b) certified occurrences, and (c) Hamming families. Solid orange lines show the across-instance means; median gains are reported in Table S5. The single positive-gain instance in each panel is highlighted in blue, and header annotations give the win/loss/tie counts.

## S5.9 Connected ER topology-transfer control

Ten connected Erdős–Rényi graphs with $N = 150$ were generated at $p = 4/(N-1) = 4/149$, targeting an unconditional mean degree of four. Disconnected draws were rejected through deterministic seed increments, and the accepted seeds were retained. The accepted graphs contained 297–346 edges (mean 317.9), compared with 296 edges for each NetworkX BA $m = 2$, $N = 150$ graph. QA-Base and DC-OSQA were evaluated on the same graph instances with 1,000 reads per protocol. Each graph was reconstructed from its released seed, its edge count was checked against the registry, and vertex-cover feasibility was recomputed directly from every raw bitstring. Certified candidates were then selected by the same feasibility-plus-$K_{\mathrm{exact}}$ rule, and complete-linkage families used $\tau = 0.1N = 15$.

The sparse-BA direction did not transfer to this small ER set. QA-Base averaged 19.9 unique certified candidates and DC-OSQA averaged 3.5, giving a mean paired gain of $-16.4$ and a median gain of $-13.5$ (one DC win, seven QA wins, and two ties; two-sided Wilcoxon $p = 2.34 \times 10^{-2}$). Certified occurrences averaged 38.4 and 4.0, respectively (mean gain $-34.4$, $p = 1.56 \times 10^{-2}$). Hamming-family counts averaged 2.3 and 1.3 (mean gain $-1.0$; one DC win, six QA wins, and three ties; $p = 7.81 \times 10^{-2}$). The comparison lies in a low-yield regime, and the connected ER graphs were matched to BA only by nominal mean degree rather than by realized edge count or degree-distribution shape. It therefore serves as an ensemble-transfer boundary; identifying the responsible structural property requires a separately matched design (Fig. S10).

Together, the regenerated-embedding, shuffled-assignment, inclusion-shift, offset-scale, density, 3-regular, and ER analyses support an empirical association between degree-aligned placement and the observed support shift in sparse BA MVC instances. The mapping

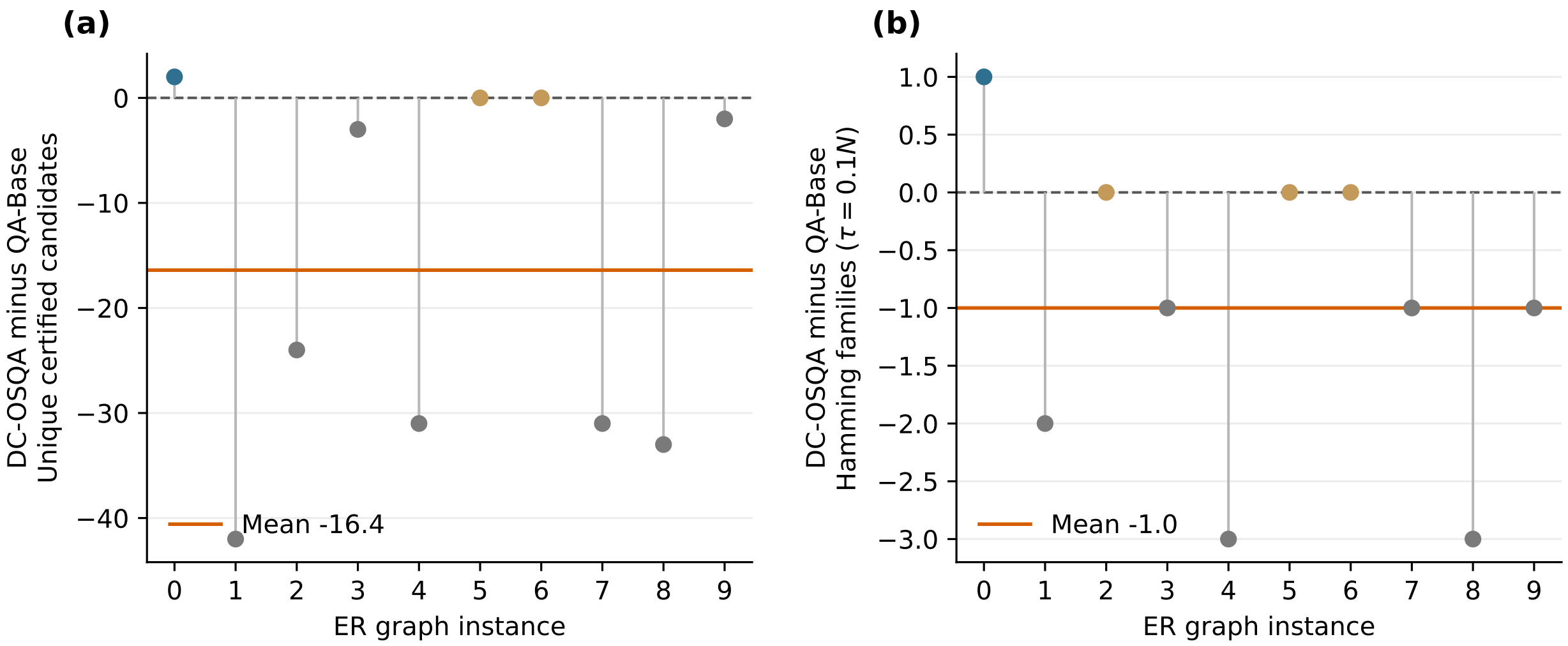


Figure S10: Connected ER topology-transfer control at $N = 150$, generated with nominal mean degree four. Points show instance-paired DC-OSQA minus QA-Base gains in (a) unique certified candidates and (b) complete-linkage Hamming families at $\tau = 0.1N$. Horizontal orange lines mark the across-instance means; negative values favor QA-Base. Blue marks the single DC-OSQA win and tan marks ties.

diagnostic argues against a solely additive monotone chain-length account, while the 3-regular and ER controls delimit transfer beyond the sparse BA ensemble.

# S6 QPU read-budget extension and robustness diagnostics

## S6.1 Direct 50,000-read comparison

The $N = 150$ extension combines an initial 10,000-read dataset with 40,000 additional reads for each of 20 matched BA $m = 2$ instances and each protocol, giving 2,000,000 QPU reads. The fixed top-10-candidate share in Table S6 is the occurrence mass of the 10 most frequent observed certified candidates, or all candidates if fewer than 10 are observed. It is distinct from the scale-normalized top-decile share in Table S1.

Table S6: Direct 50,000-read QPU comparison. W/L denotes DC-OSQA wins/QA-Base wins; for concentration metrics, the favorable direction is stated explicitly. Mean DC–QA is the paired mean difference, and $p$-values are from two-sided paired Wilcoxon tests.

| Metric | DC mean | QA mean | W/L | Mean DC–QA | $p$ |
|---|---|---|---|---|---|
| Unique certified | 15929.6 | 2812.2 | 19/1 | 13117.4 | $3.81 \times 10^{-6}$ |
| Certified occurrences | 31976.2 | 21528.8 | 16/4 | 10447.5 | $1.36 \times 10^{-2}$ |
| Normalized entropy (higher) | 0.94 | 0.81 | 19/1 | 0.13 | $3.81 \times 10^{-6}$ |
| Simpson support fraction (higher) | 0.54 | 0.15 | 19/1 | 0.38 | $2.67 \times 10^{-5}$ |
| Top-10-candidate share (lower) | 0.04 | 0.26 | 20/0 | −0.22 | $1.91 \times 10^{-6}$ |
| Gini coefficient (lower) | 0.36 | 0.65 | 18/2 | −0.28 | $1.34 \times 10^{-4}$ |
| Duplicate rate (lower) | 0.46 | 0.73 | 18/2 | −0.27 | $5.86 \times 10^{-4}$ |

## S6.2 Yield–diversity separation and threshold sensitivity

Figure S11 jointly reports certified yield, conditional rarefaction, Hamming-threshold sensitivity, and union-support complementarity.

Certified yield is

$$Y = \frac{C}{R}, \tag{S4}$$

where $C$ is the number of certified occurrences among $R$ raw reads. Mean yield at 50,000 reads was 0.64 for DC-OSQA and 0.43 for QA-Base.

To compare diversity after conditioning on the same number $m$ of certified draws, exact finite-population rarefaction [8] was computed from occurrence counts $n_j$:

$$\mathbb{E}[U_m] = \sum_{j=1}^{M} \left[ 1 - \frac{\binom{C-n_j}{m}}{\binom{C}{m}} \right], \qquad 0 \le m \le C. \tag{S5}$$

At $m = 1000$, 15 graphs had at least 1000 certified occurrences under both protocols; mean expected support was 833.1 for DC-OSQA and 449.5 for QA-Base. At $m = 10{,}000$, 13 paired-eligible graphs remained, with mean expected support 5707.8 for DC-OSQA and 1870.7 for QA-Base. Higher expected support under DC-OSQA therefore remained after conditioning on equal numbers of certified draws.

The complete-linkage analysis was repeated for $\tau/N \in \{0, 0.05, 0.10, 0.15, 0.20\}$, retaining zero-yield groups as zero. Mean paired family gains (DC-OSQA minus QA-Base) were

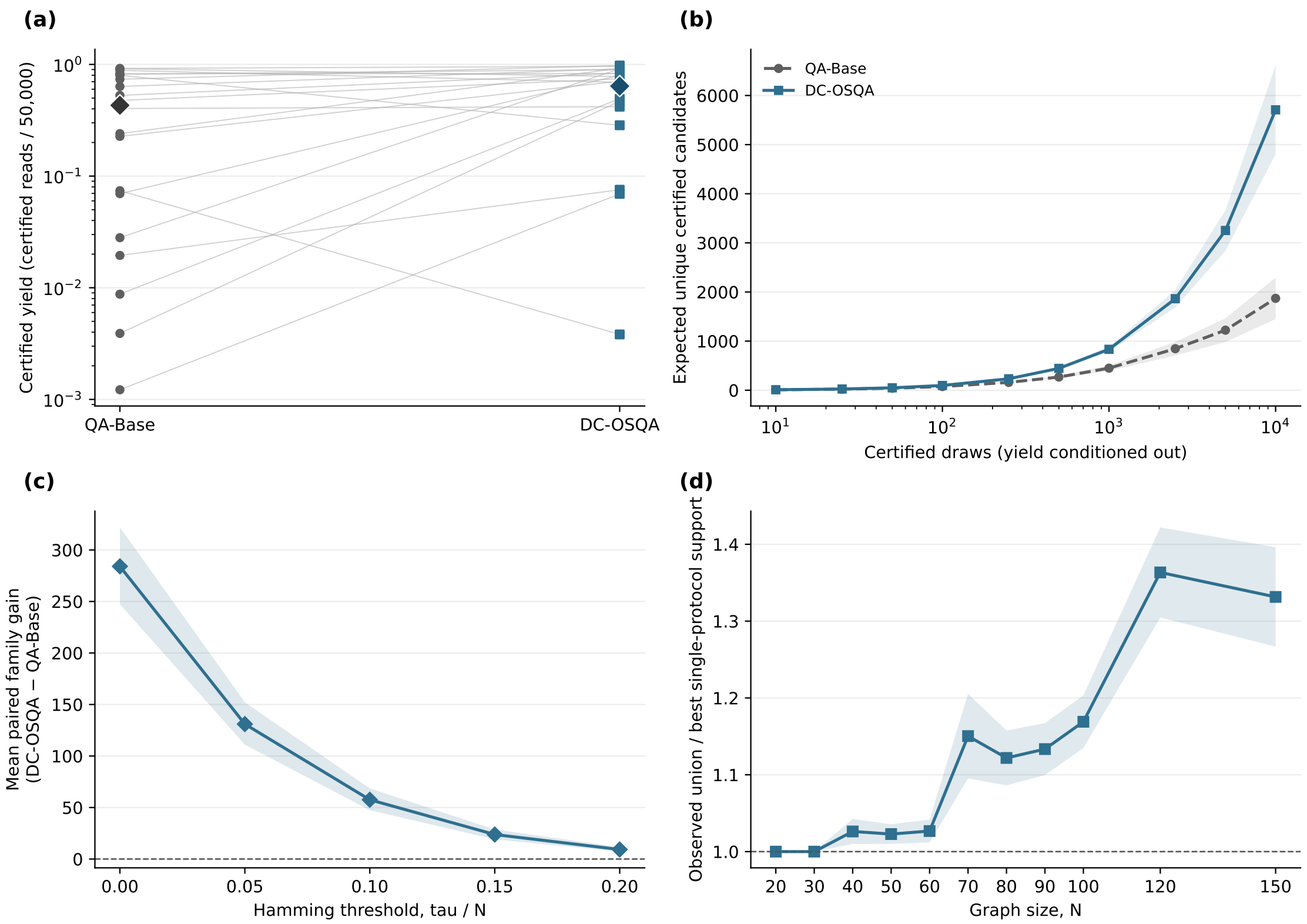


Figure S11: Yield–diversity separation and robustness. (a) Certified yield at 50,000 reads; gray lines connect matched instances and diamonds mark protocol means. (b) Exact rarefaction conditional on the same number of certified draws. Lines show means and bands show s.e.m. across instances eligible for both protocols at each depth ($n = 20$ at $m = 10$ and $n = 13$ at $m = 10{,}000$). (c) Mean paired Hamming-family gain versus normalized threshold; the shaded band is the graph-size-stratified bootstrap 95% percentile confidence interval. (d) Observed union support divided by the larger single-protocol support; the band shows s.e.m. across graph instances, and the dashed line marks equality.

284.08, 130.98, 57.56, 23.79, and 9.25; stratified-bootstrap 95% intervals remained above zero throughout. For both the threshold analysis and the primary effect sizes, 5,000 bootstrap samples were generated by resampling graph-level pairs with replacement within graph-size strata; the 2.5th and 97.5th percentiles define the reported 95% intervals. Cohen's paired effect was $d_z = \overline{\Delta}/s_\Delta$, using the graph-level paired differences.

Finally, $|S_{\mathrm{QA}} \cup S_{\mathrm{DC}}|$ is a lower bound on optimum degeneracy. Every tested graph had at least two observed certified optima. Mean union-to-larger-support ratios for $N = 70, 80, 90, 100, 120, 150$ were 1.15, 1.12, 1.13, 1.17, 1.36, and 1.33, respectively; these quantify complementarity within the observed certified support.

## S6.3 Matched-budget support recovery

The matched-budget overlap summaries are reported in Table S7.

Table S7: Support recovery and overlap at $N = 150$. Shared is the mean intersection size, recovery is the fraction of the target observed support found by the source, and union gain is the union size divided by the larger individual support.

| Source | Target | Shared | Target recovered | Jaccard | Union gain |
|---|---|---|---|---|---|
| QA-Base 50k | DC-OSQA 50k | 740.5 | 0.09 | 0.08 | 1.09 |
| DC-OSQA 50k | QA-Base 50k | 740.5 | 0.43 | 0.08 | 1.09 |
| SA 50k | DC-OSQA 50k | 2226.3 | 0.32 | 0.14 | 1.36 |
| DC-OSQA 50k | SA 50k | 2226.3 | 0.17 | 0.14 | 1.36 |

# S7 Chain-break diagnostics

In the 10,000-read $N = 150$ comparison, no chain-break metric showed a statistically significant paired difference between protocols (Table S8). Figure S5 provides the complementary vertex-level chain-length analysis on the canonical $N = 70$–$150$ mappings.

Table S8: Chain-break diagnostics for the 10,000-read $N = 150$ QPU comparison, paired over 20 BA $m = 2$ instances. For each instance, occurrence counts weight the read-level values; the table reports the across-instance means of those statistics as percentages. "Fraction over 1%" is the percentage of reads whose chain-break fraction exceeds 0.01. Differences are DC-OSQA minus QA-Base in percentage points, and $p$-values are from two-sided paired Wilcoxon tests.

| Metric | DC mean (%) | QA mean (%) | Mean diff. | Median diff. | $p$ |
|---|---|---|---|---|---|
| Mean fraction | 0.10 | 0.06 | 0.04 | 0.00 | 0.11 |
| Maximum fraction | 0.60 | 0.60 | 0.00 | 0.00 | 1.00 |
| Fraction over 1% | 4.52 | 3.35 | 1.16 | 0.00 | 0.75 |

# S8 Idealized closed-system proof of principle

## S8.1 Question and model

We used a separate small-system calculation to test a sufficiency claim: whether a degree-aligned local annealing schedule can, by itself, redistribute occupation within a degenerate MVC optimum manifold and increase expected distinct-optimum discovery. The calculation is an idealized unitary QA model rather than a simulation of the QPU. Its role follows earlier small-system studies that used direct Schrödinger evolution or perturbative analysis to examine biased ground-state sampling and nonuniform drivers [9, 10, 11].

For each logical BA $m = 2$ graph, the dimensionless time-dependent Hamiltonian was

$$H(s) = -\sum_i A(s_i)\sigma_i^x + \sum_i B(s_i)h_i\sigma_i^z + \sum_{(i,j)\in E} J_{ij}\sqrt{B(s_i)B(s_j)}\,\sigma_i^z\sigma_j^z, \tag{S6}$$

where the MVC QUBO weights were $\lambda_{\text{card}} = 1$ for cardinality and $\lambda_{\text{edge}} = 2$ for an uncovered-edge penalty, giving $h_i = 1/2 - d_i/2$ and $J_{ij} = 1/2$. The primary envelopes were $A(u) = 1 - u$ and $B(u) = u$. A local offset defined

$$s_i(s) = \text{clip}[s + 4\delta_i s(1-s), 0, 1]\,, \qquad \delta_i = -\alpha\frac{d_i}{d_{\max}}, \tag{S7}$$

which preserves the common endpoints and remains monotone over the tested range $\alpha \le 0.2$. The *zero-offset baseline* set $\delta_i = 0$ for every vertex, so that $s_i(s) = s$. The *mean-matched uniform-offset control* instead assigned every vertex $-\alpha\,\overline{d/d_{\max}}$, the mean of the degree-aligned offsets. The square-root factor in the coupler term is the symmetric geometric-mean extension of the local problem envelopes.

The state was initialized as $|+\rangle^{\otimes N}$ and evolved for dimensionless duration $T = 40$ with 600 propagation steps. Writing the global schedule fraction as $s = t/T$, each step had duration $\Delta t = T/600$ and was evaluated at $\bar{s} = (t + \Delta t/2)/T$. Each step used the symmetric second-order Suzuki–Trotter (Strang) splitting [12, 13]

$$U(t, t+\Delta t) \simeq e^{-\mathrm{i}H_{\mathrm{P}}(\bar{s})\Delta t/2}\, e^{-\mathrm{i}H_{\mathrm{D}}(\bar{s})\Delta t}\, e^{-\mathrm{i}H_{\mathrm{P}}(\bar{s})\Delta t/2}, \tag{S8}$$

Here $H_{\mathrm{D}} = -\sum_i A(s_i)\sigma_i^x$ is a sum of mutually commuting single-qubit terms and is therefore exponentiated exactly as a product of independent rotations, whereas $H_{\mathrm{P}}$ collects the longitudinal and coupler terms of Eq. (S6), which are diagonal in the computational basis and applied as an exact phase. The driver and problem factors are evaluated without further approximation at the step midpoint; together with midpoint evaluation of the time dependence, the scheme has local error $O(\Delta t^3)$ and global error $O(\Delta t^2)$. For the convergence check, $\Delta S_{\mathrm{eff}} = S_{\mathrm{eff}}^{\mathrm{degree}} - S_{\mathrm{eff}}^{\mathrm{baseline}}$, where $S_{\mathrm{eff}}$ is defined below. The check used the $N = 12$, seed 12043 instance at $\alpha = 0.1$ under the linear envelopes. Increasing the step count from 300 to 1200 changed $\Delta S_{\mathrm{eff}}$ by 0.6%, and the change from 600 to 1200 steps was below 0.2%. Exact enumeration on the $N = 12$ instances with seeds 12042–12044 verified that the terminal Ising ground manifold equaled the minimum-cover manifold.

For every graph, all $2^N$ bitstrings were enumerated exhaustively; the feasible covers of minimum cardinality formed $\mathcal{M}$, and $|\mathcal{M}|$ gave its degeneracy. For each $N \in \{12, 14, 16, 18, 20\}$, we scanned the fixed seeds $42 + 1000N + i$, $i = 0, \ldots, 59$, and retained every graph with at least four exact minimum covers. This gave 32, 32, 37, 32, and 48 graphs, respectively. The degeneracy threshold defines the population in which a diversity change is measurable and was applied without reference to the simulated effect. The primary calculation used $\alpha = 0.1$. At all five sizes, assignment controls compared the degree-aligned rule with the mean-matched uniform offset, 20 random permutations of the same degree-aligned offset multiset, a degree-reversed assignment that placed the most negative values on the lowest-degree vertices, and 20 degree-independent assignments drawn independently from Uniform$[-\alpha, 0]$. The degree-aligned and shuffled conditions were also evaluated under linear, convex-power, and trigonometric envelope families at all five sizes, defined respectively by $(A(u), B(u)) = (1-u, u)$, $((1-u)^{1.6}, u^{1.8})$, and $(\cos(\pi u/2), \sin(\pi u/2))$. All three families shared $A(0) = B(1) = 1$ and $A(1) = B(0) = 0$, so this comparison changed the path shape without changing its endpoints.

For $x \in \mathcal{M}$, let $w_x = |\langle x|\psi(T)\rangle|^2$. The target-manifold occupation probability was $Y_{\mathcal{M}} = \sum_{x\in\mathcal{M}} w_x$, and the conditional probabilities were $p_x = w_x/Y_{\mathcal{M}}$. Conditional effective support was

$$S_{\mathrm{eff}} = \left(\sum_{x\in\mathcal{M}} p_x^2\right)^{-1}. \tag{S9}$$

Unlike the empirical Simpson effective-support fraction, which divides the corresponding effective support by the observed support size $M$, $S_{\mathrm{eff}}$ is an unnormalized effective number

of optimum states. Expected unconditional discovery under $R$ independent draws from the final Born distribution was

$$\mathbb{E}[U_R] = \sum_{x\in\mathcal{M}} \left[1 - (1 - w_x)^R\right]. \tag{S10}$$

The fixed discovery-budget grid was obtained by rounding $10^{k/8}$ to the nearest integer for $k = 0, \ldots, 32$ and removing duplicates, yielding 31 budgets from $R = 1$ to $10^4$. Paired mean contrasts used 5,000 graph-level bootstrap resamples, with 95% intervals taken as the 2.5th and 97.5th percentiles of the resampled statistic. For the discovery curves, a two-sided maximum-absolute test used 5,000 graph-level sign-flip permutations over this grid, applying one sign to each graph's complete curve. The Monte Carlo $p$-value was $p = (1 + n_{\geq})/(5000 + 1)$, where $n_{\geq}$ counts permuted statistics at least as large as the observed statistic; the five size-specific $p$-values were then Bonferroni adjusted.

## S8.2 Endpoint redistribution and assignment controls

At the fixed $\alpha = 0.1$ operating point, mean degree-aligned-minus-zero-offset effective-support gains were 0.19, 0.19, 0.14, 0.19, and 0.24 for $N = 12, 14, 16, 18$, and 20, respectively. The corresponding degree-aligned-minus-shuffled gains were 0.29, 0.34, 0.31, 0.42, and 0.45, with positive bootstrap intervals at all five sizes; degree-aligned support exceeded the within-graph shuffle mean in 31/32, 29/32, 28/37, 30/32, and 43/48 graphs. Discovery gains in $\mathbb{E}[U_R]$ were positive over an intermediate budget range at all five sizes, with the mean curves reaching their descriptive maxima at $R \approx$ 32–100. Curve-wise inference used the two-sided maximum-absolute sign-flip test, which accounts for selection of the most extreme budget within a curve; after Bonferroni adjustment across the five sizes, the $p$-values were $1.00 \times 10^{-3}$, $2.10 \times 10^{-2}$, $2.00 \times 10^{-3}$, 0.21, and $2.20 \times 10^{-2}$. The curve-wise test therefore survived correction at four of the five sizes, with $N = 18$ remaining positive descriptively but not significant after correction.

The mean effective-support change was positive for both degree-aligned and mean-matched uniform assignments at all five sizes, with the degree-aligned mean larger in each case. Shuffled, degree-reversed, and random assignments instead had negative mean changes at every size. The matched-shuffle contrast therefore provided the clearest placement control. Its direction was retained for every tested envelope family and size, and the degree-aligned value exceeded the within-graph shuffle mean in a majority of graphs in all 15 size–envelope combinations (Fig. S12).

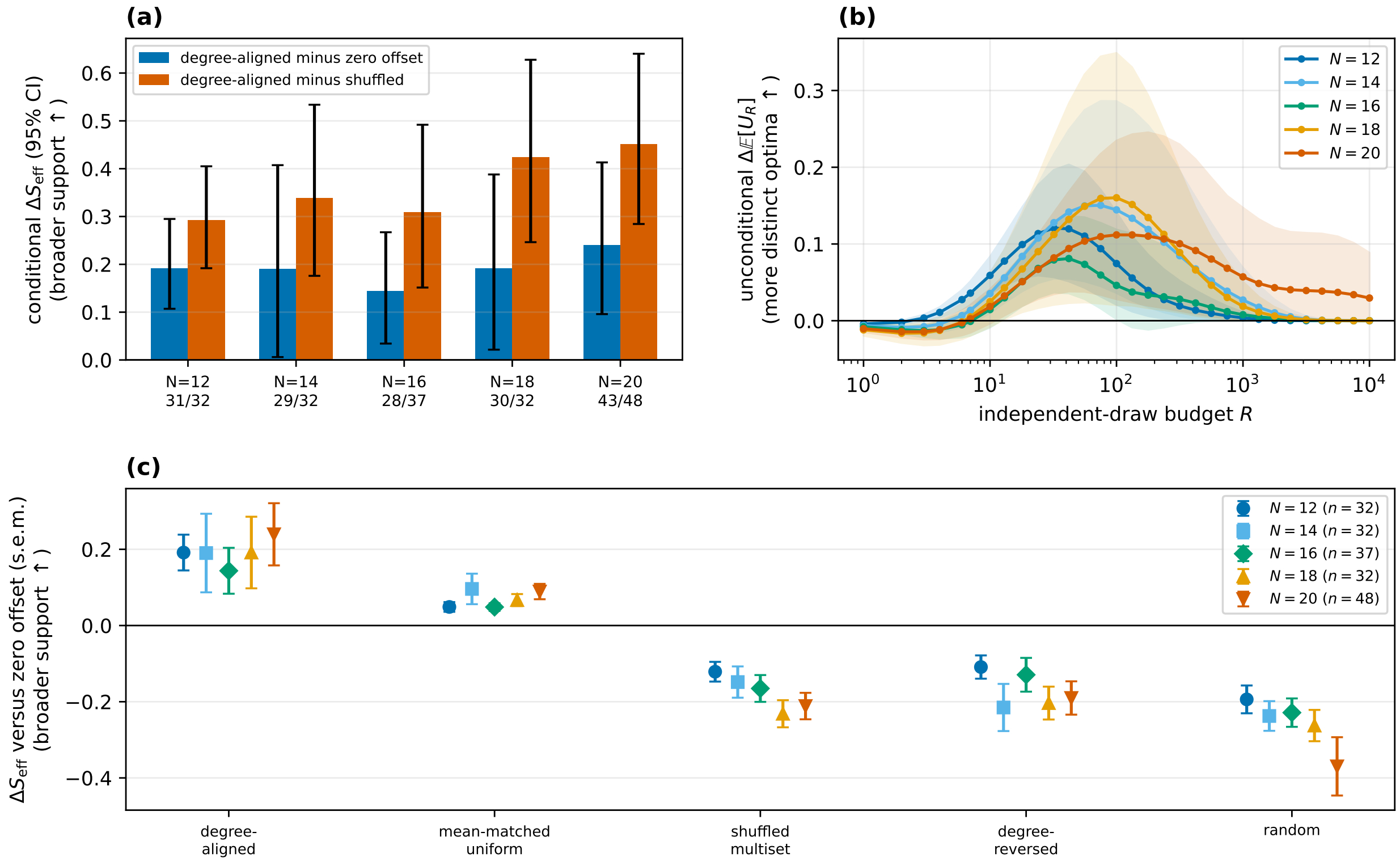


Figure S12: Endpoint proof of principle in the idealized closed-system model at $\alpha = 0.1$. (a) Conditional effective-support contrasts across $N = 12$–$20$; bars show paired means and 95% bootstrap intervals, and each fraction below the size gives the number of graphs in which the degree-aligned value exceeded the within-graph mean of 20 shuffled assignments. (b) Degree-aligned minus zero-offset expected distinct-optimum discovery versus independent-draw budget $R$; bands are pointwise 95% bootstrap intervals. The two-sided maximum-absolute sign-flip $p$-values were adjusted across the five sizes. (c) Assignment controls across all five sizes; points are across-graph means and error bars are s.e.m.

## S8.3 Operating window and time-resolved redistribution

All five size pools were evaluated on the fixed sensitivity grid $\alpha \in \{0.05, 0.10, 0.15, 0.20\}$. At $\alpha = 0.05$ and 0.10, the effective-support gains had positive bootstrap intervals at every size. Increasing $\alpha$ made the target-manifold occupation shift progressively more negative. At $\alpha = 0.15$, the support interval remained positive at $N = 12$ and 20 but included zero at $N = 14, 16$, and 18; at $\alpha = 0.20$, only $N = 12$ retained a positive interval. These per-size intervals are not adjusted for multiplicity across the grid. The sensitivity analysis therefore supports $\alpha = 0.10$ as the most consistent tested operating point and identifies a broadening–occupation trade-off rather than a universal optimal offset scale (Fig. S13).

To locate when the contrast formed within the model, the evolving state was recorded at 20 fixed anneal fractions, $s \in \{0.05, 0.10, \ldots, 1.00\}$. At intermediate $s$,

$$Y_{\mathcal{M}}(s) = \sum_{x \in \mathcal{M}} |\langle x|\psi(s)\rangle|^2 \tag{S11}$$

is the projection onto the *final* exact MVC manifold, not an instantaneous-ground-state population. Degree alignment reduced this target-manifold occupation relative to the zero-offset baseline through most of the evolution, with final mean occupation shifts ranging

from −0.41 to −1.20 percentage points across the five sizes at $\alpha = 0.1$. In contrast, its conditional effective-support advantage developed across the anneal and remained positive at every endpoint. Unlike the QPU data, in which certified yield also increased, the coherent surrogate produced a positive discovery contrast without an occupation increase: the broader conditional distribution was sufficient to outweigh the occupation penalty over the intermediate budget range. Panels (c) and (d) of Fig. S13 therefore separate the two dynamical components: the occupation penalty and the conditional broadening form over the same evolution but in opposite directions. The time-resolved calculation does not identify a device freeze-out point.

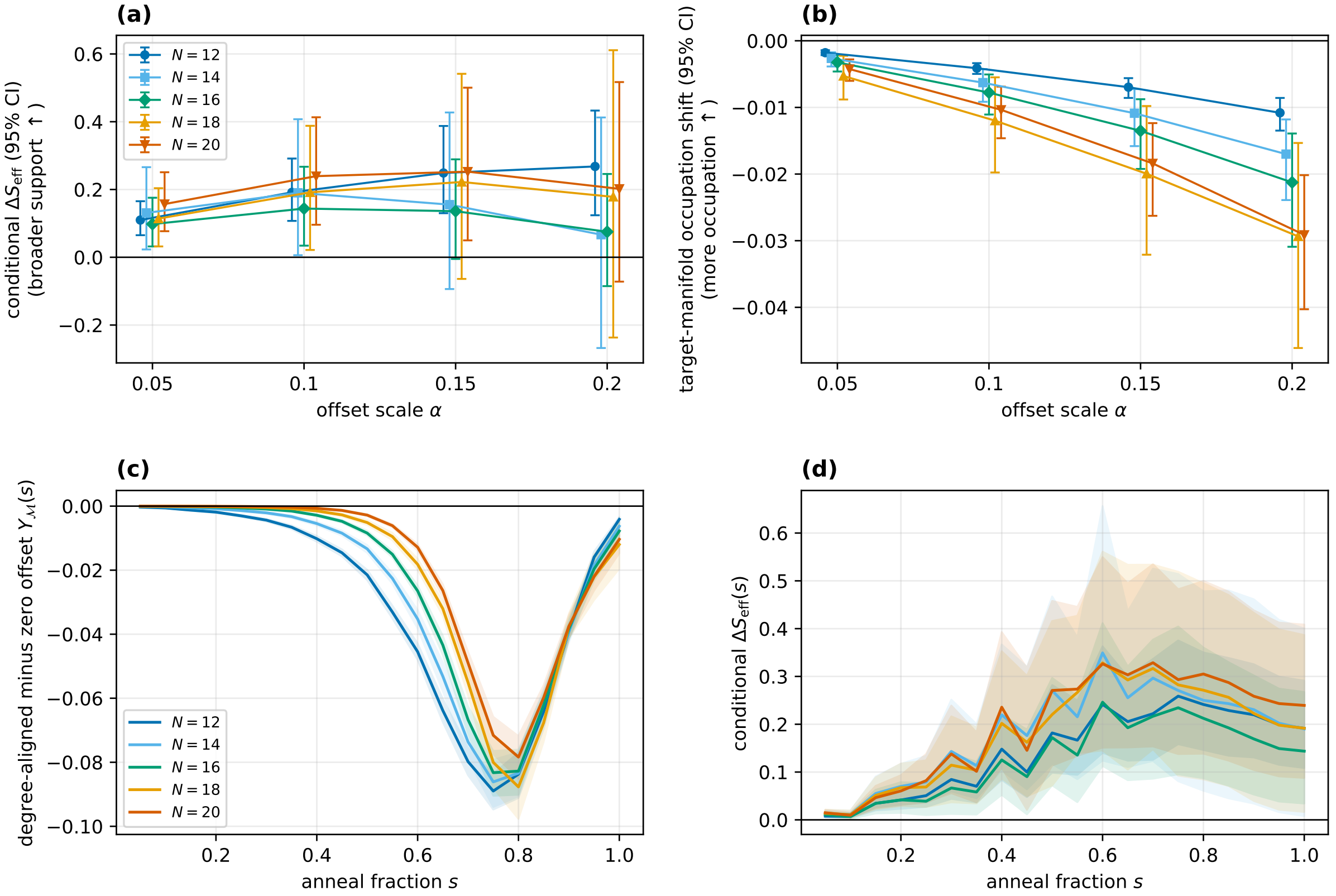


Figure S13: Operating-window and time-resolved diagnostics across $N = 12$–$20$. (a,b) Degree-aligned minus zero-offset changes over the fixed $\alpha$ grid in conditional effective support and target-manifold occupation; points are paired means and error bars are 95% bootstrap intervals. (c) Degree-aligned minus zero-offset change in probability projected onto the final exact MVC manifold during the evolution. (d) Mean conditional effective-support contrast during the same evolution. Bands in (c,d) are pointwise 95% bootstrap intervals across the eligible graphs at each size and are not simultaneous over the anneal fraction.

Taken together, these calculations establish a model-level sufficiency result: degree-aligned local schedules can broaden occupation within a degenerate MVC optimum manifold and increase expected finite-budget discovery under the specified finite-time coherent model over an intermediate budget range. The empirical QPU and idealized calculations independently show sensitivity to logical-variable placement relative to shuffled assignments.



\end{document}